\documentclass[11pt]{article}

\usepackage[final]{acl}

\usepackage{times}
\usepackage{latexsym}
\usepackage{xurl}
\usepackage{times}
 \usepackage{CJKutf8}
\usepackage{bm}
\newtheorem{definition}{Definition}
\usepackage{hyperref}
\usepackage{xcolor}
\usepackage{lineno}
\usepackage{verbatim}
\usepackage{booktabs} 
\usepackage{algorithm}
\usepackage{algorithmic}
\usepackage{epsfig}
\usepackage{color}
\usepackage{amsmath}
\usepackage{graphicx,subfigure}
\usepackage{multirow}
\usepackage{makecell}
\newcommand{\re}[1]{\textcolor{black}{#1}}
\usepackage{amssymb}
\usepackage[most]{tcolorbox} % 引入tcolorbox宏包
\tcbuselibrary{theorems} % 引入theorems库
\usepackage{helvet}
\usepackage{threeparttablex}
\newtcolorbox{system}[1]{
  colback=blue!5,
  colframe=blue!35!black,
  fonttitle=\bfseries,
  title={System-Mode Self-Reminder},
  }
\newtcolorbox{response}[1]{
  colback=green!5,
  colframe=green!35!black,
  fonttitle=\bfseries,
  title={ChatGPT Defended by Self-Reminder},
  }

  \newtcolorbox{new-system}[1]{
  colback=blue!5,
  colframe=blue!35!black,
  fonttitle=\bfseries,
  }

\newtcolorbox{query}[1]{
  colback=blue!5,
  colframe=blue!35!black,
  fonttitle=\bfseries,
  title={User Query},
  }
\newtcolorbox{response-self}[1]{
  colback=green!5,
  colframe=green!35!black,
  fonttitle=\bfseries,
  title={ChatGPT Defended by Self-Reminder},
  }

\newtcolorbox{new-response}[1]{
  colback=brown!5,
  colframe=brown!35!black,
  fonttitle=\bfseries,
  title={ChatGPT},
  }
\newcommand{\new}[1]{\textcolor{black}{#1}}
  \usepackage{subfigure}% For proper rendering and hyphenation of words containing Latin characters (including in bib files)
\usepackage[T1]{fontenc}
\usepackage[utf8]{inputenc}

\usepackage{microtype}

\usepackage{inconsolata}

\usepackage{graphicx}

\title{Measuring Human Contribution in AI-Assisted Content Generation}

\author{
\textbf{Yueqi Xie\textsuperscript{1}\thanks{Joint First Authors}\thanks{Corresponding Authors}},
\textbf{Tao Qi\textsuperscript{2}\footnotemark[1]\footnotemark[2]},
\textbf{Jingwei Yi\textsuperscript{3}\footnotemark[1]\footnotemark[2]},
\textbf{Xiyuan Yang\textsuperscript{4}},
\textbf{Ryan Whalen\textsuperscript{5}},
\\
\textbf{Junming Huang\textsuperscript{1}},
\textbf{Qian Ding\textsuperscript{6}},
\textbf{Yu Xie\textsuperscript{1,7}},
\textbf{Xing Xie\textsuperscript{6}},
\textbf{Fangzhao Wu\textsuperscript{6}\footnotemark[2]}
\\
\\
\textsuperscript{1}Princeton University,
\textsuperscript{2}Tsinghua University,\\
\textsuperscript{3}University of Science and Technology of China,
\textsuperscript{4}University of Illinois Urbana-Champaign,
\\
\textsuperscript{5}The University of Hong Kong,
\textsuperscript{6}Microsoft Research Asia,
\textsuperscript{7}Peking University
\\
\small{
\textbf{Correspondence:}
\href{mailto:yueqixie@princeton.edu}{yueqixie@princeton.edu},
\href{mailto:taoqi.qt@gmail.com}{taoqi.qt@gmail.com},
\href{mailto:jwyi1029@gmail.com}{jwyi1029@gmail.com},
\href{mailto:fangzwu@microsoft.com}{fangzwu@microsoft.com}
}
}
\begin{document}
\maketitle
\begin{abstract}
With the growing prevalence of generative AI, an increasing amount of content is no longer exclusively generated by humans but by generative AI models with human guidance. This shift presents notable challenges for the delineation of originality due to the varying degrees of human contribution in AI-assisted works.
This study raises the research question of measuring human contribution in AI-assisted content generation and introduces a framework to address this question that is grounded in information theory. By calculating mutual information between human input and AI-assisted output relative to self-information of AI-assisted output, we quantify the proportional information contribution of humans in content generation. Our experimental results demonstrate that the proposed measure effectively discriminates between varying degrees of human contribution across multiple creative domains. To further enhance real-world applicability, we extend the framework to estimate the minimal necessary human contribution for any text without requiring human input and validate its effectiveness. We hope that this work lays a foundation for measuring human contributions in AI-assisted content generation in the era of generative AI.
% \footnote{The source code is available at \url{https://anonymous.4open.science/r/Human-Contribution-Measurement-44BF/}.}
\end{abstract}

%\section{Introduction}

%\section{Introduction}
\begin{figure*}[t!]
    \centering
    \includegraphics[width=1.0\linewidth]{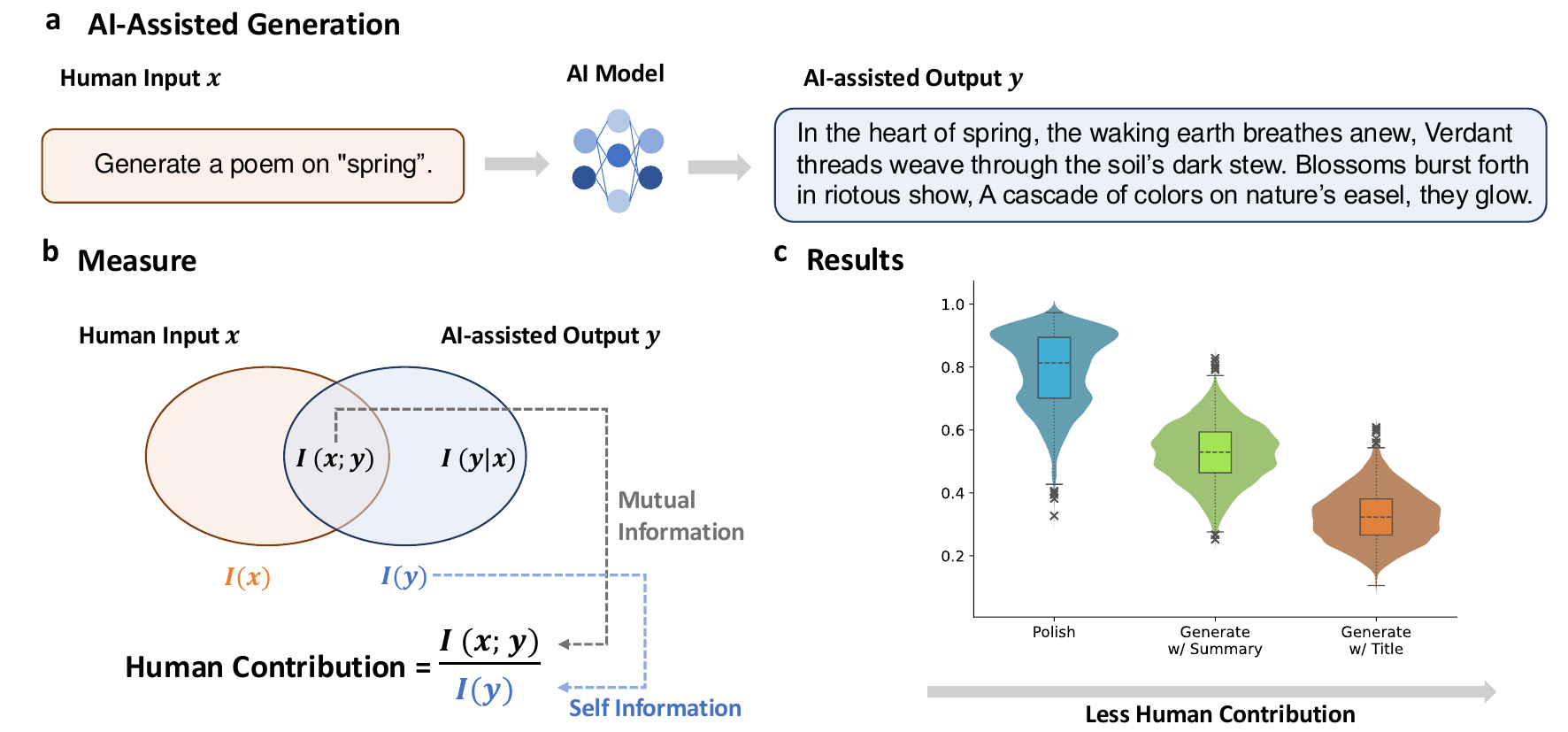}
    \caption{
    \textbf{a}. Illustration of AI-assisted content generation, where an AI model is prompted with human input and generates output.
    \textbf{b}. Overview of the proposed method for measuring human contribution, quantified by the ratio of mutual information between human input and AI-assisted output to the total self-information of the AI-assisted output.
    \textbf{c}. Outcomes of our proposed measure across various poem generation scenarios using Llama-3, involving varying degrees of human contribution (polishing a human poem, generation with the summary, in other words, key ideas, of a human poem, and generation with a poem title). 
} 
    \label{fig:overview}
\end{figure*}
% \textbf{d},generally demonstrates lower values for AI-generated content with less human contribution, 
%     divergence between the estimated AI-assisted distribution, and the generative distribution of the AI model when not conditioned on user input, averaged across tokens.
     % The proposed measurement can be applied to copyright-related law development and online regulation.}
% contribution evaluation in various scenarios.  Although all involve human-AI collaboration, the distinctions in human contribution are expected to be substantial. 
% Illustration of our proposed evaluation measurement distribution across different datasets and scenarios when ChatGPT is applied for generation and Llama is applied for evaluation.  It can be observed that the proposed measurement measurement is capable of ideally distinguishing various levels of human contribution
% and revising the AI-generated content, collaborating with AI to ultimately produce the final creation.
\section{Introduction}
Recent advances in large language models (LLMs) have impacted our personal and working lives in significant ways, notably by changing the process of content generation~\cite{wingstrom2024redefining}.
AI ``copilots'' have emerged as a new and powerful content production tool across a variety of domains, such as lyrics creation~\cite{zhang2022youling}, office work~\cite{office,zheng2022telling},  academic writing~\cite{dergaa2023human}, etc. 
% In fact, experimental evidence has shown that the productivity gains with generative AI have been significant~\cite{liu2023experimental}. 
Consequently, an increasing amount of new content being generated is no longer solely created by humans but is rather the result of AI-assisted creation~\cite{fui2023generative,wang2020human,10.1145/3581641.3584052}. 
In this new creative modality, humans contribute by providing prompts to AI models, resulting in the generation of ``AI-assisted output", as illustrated in Figure~\ref{fig:overview}\textbf{a}.

This development has raised debates about determining the originality and corresponding regulation of content generated with AI assistance~\cite{AI-Copyright,fui2023generative}.
The varying degrees of human contribution in AI-assisted generation complicate the attribution of intellectual contribution to AI-assisted outputs. This issue is particularly pertinent in fields that prioritize originality, such as education~\cite{hutson2024rethinking}, academic research~\cite{yu2023reflection,nakadai2023ai,kwonai}, and creative work~\cite{us_copyright_office_2023_ai}. 
For example, universities face a dilemma in whether to ban or embrace AI. Administrators and instructors are concerned that students might use AI to create materials for evaluation with varying levels of originality, potentially compromising educational fairness and effectiveness~\cite{VOA2023,NYT2023}.
Similarly, there is a growing debate, underscored by notable incidents~\cite{us_copyright_office_2022,us_copyright_office_2023}, concerning the copyright eligibility of AI-assisted works~\cite{us_copyright_office_2023_ai, abbott2023disrupting, hristov2016artificial}. 

At the two extreme ends of the human–AI contribution spectrum, the attribution of originality is relatively clear. If a human author simply uses AI to polish their document, it should be considered the result of the author's own work. Conversely, if a human uses a short, less-informative prompt to generate a large amount of text, it will not reflect much of the human's intellectual conception. 
However, there remains a substantial grey area between these two extremes, in which determining originality requires insight into the \emph{degree of human contribution} during the AI-assisted generation process.
Hence, there is an urgent need for a credible measure by which to evaluate human contribution in AI-assisted content generation.

In this paper, we address the quantification of human contribution in AI-assisted content generation. We begin with the recognition that a major obstacle is the lack of a well-defined perspective, or medium, by which to ascertain the extent to which content output can be attributed to humans rather than the AI tools they have used. Towards this goal, we introduce a new general framework within which we provide a preliminary attempt to quantify human contribution in AI-assisted content generation. 
Our framework hinges on the concept of \emph{information content} as a modeling medium. Utilizing principles from information theory~\cite{shannon1948mathematical}, as depicted in Figure~\ref{fig:overview}\textbf{b}, our approach centers on the quantification of the proportion of the information content in the AI-assisted output that can be attributed to human input.
% Specifically, the total information content (surprisal) in the AI-assisted output is calculated using the probability of generating the AI-assisted content without human input, which we refer to as \emph{self-information}. Meanwhile, human input introduces additional information gain in the generation process, which is calculated by the reduction in surprisal when generating the AI-assisted content with human input compared to without it.
% We define this information gain as the \emph{mutual information} between the human input and AI-assisted output. 
% The ratio of mutual information to self-information is then used as our measure of human contribution. This measure quantifies the distinct informational contributions of the human relative to the total information in the generation process.
Specifically, it is a ratio of two quantities.  The denominator is 
the total/unconditional information content (surprisal) in AI-assisted output, calculated as the negative logarithm of the probability of generating the AI-assisted generated content, which we refer to as \emph{self-information}, {$I(\boldsymbol{y})$}. 
The numerator,  {$I(\boldsymbol{x};\boldsymbol{y})$}, is the portion of self-information {$I(\boldsymbol{y})$} that is shared with the total information content from human input, {$I(\boldsymbol{x})$}, which we define as \emph{mutual information}. 
The difference between the two is the \emph{conditional self-information} in AI-assisted output given user input, {$I(\boldsymbol{y}|\boldsymbol{x})$}, calculated as the negative logarithm of the probability of generating the AI-assisted output conditional on human input. 

\begin{table*}[t!]
\centering
\caption{Detailed statistics of the constructed dataset.}
\scalebox{0.85}{
\begin{tabular}{cccccc}
\Xhline{1.0pt}
Type   & Corpus         
  &\makecell{\# Content Words} &  \makecell{\# Summary Words}&\makecell{\# Title Words} &\makecell{\# Subject Words} \\ \midrule
Paper Abstract & Arxiv & 134.24±63.07 &68.38±20.59& 9.63±3.79& 2.87±1.37 \\
% & OpenReview  & 20 & 2000 & 9.48 & 194.23 \\
% & Nature \\
% \midrule
\multirow{1}{*}{News} &  News Articles &532.85±86.26& 78.36±16.76 & 8.91±2.32 &4.06±1.00\\
% \midrule
\multirow{1}{*}{Patent Abstract} &  HUPD & 171.65±24.22 & 59.89±13.85& 8.57±5.28 & 3.91±0.79\\
% \multirow{1}{*}{Patent Discription} &  Google Patent & 100 & \\
\multirow{1}{*}{Poem} & Poetry Foundation  &208.18±94.08&48.20±11.27& 3.65±2.80 & -\\
% & Sina& \\
% & & \\
% \midrule
% Patent  Abstract & \\
% Novel  &     \\ 
\Xhline{1.0pt}
\end{tabular}
}
\label{tab:stat}
\end{table*}

We systematically validate the proposed method as a reliable measure of human contribution by evaluating its effectiveness, domain adaptivity, and model adaptivity. To achieve this, we construct a comprehensive dataset of AI-assisted content generation, encompassing various levels of human contribution, multiple creative domains, and outputs from different LLMs. 
For instance, Figure~\ref{fig:overview}\textbf{c} illustrates the distribution of the outcomes of our proposed measure for AI-assisted poem generation, 
across three varying levels of human contribution, ranging from high to low, using the LLM Llama-3.
Our proposed measure effectively discriminates between varying degrees of contribution, generally producing lower values for content with less human contribution.
Additionally, we investigate the impact of content length, writing style, model temperature, resilience to adaptive attacks, and generalization of our method in evaluation.
We further apply our measure to real-world human–AI co-creation data, demonstrating its practical applicability.

To broaden the applicability of our human contribution evaluation framework to real-world scenarios, where humans may engage in complex content generation processes or where human input is unknown, we propose an extension that estimates human contribution based on any text alone, without requiring access to its generation process.
The contributions are summarized as follows:
\begin{itemize}
    \item We formulate a novel research question aimed at quantitatively evaluating informational human contribution in AI-assisted generation.
    \item We propose simple yet effective information-theoretic measures to approach this problem.
    \item We construct a dataset of AI-assisted generation with varying levels of human contribution for rigorous evaluation.
    \item We systematically evaluate the proposed methods on this dataset, examining overall effectiveness, the impact of content length, writing style,  model temperature, resilience to adaptive attacks, and generalization performance.
    \item We extend the framework to estimate the human contribution of any text alone, without requiring access to the generation process.
\end{itemize}

\section{Related Work}
The research problem most closely related to evaluating human contribution is the detection of content generated by LLMs~\cite{yang2023survey,wu2025survey}.
As the performance of LLMs continues to improve, the risk of being unable to distinguish between content generated by LLMs and humans becomes increasingly apparent, with attendant threats in security, fraud prevention~\cite{pan2023risk,xie2023defending}, and academic integrity~\cite{bin2023use}, among other fields~\cite{yang2023survey,kumar-etal-2023-language, yang2023shadow}. 
Consequently, researchers are increasingly directing their efforts towards the detection of LLM-generated content, specifically ascertaining whether a given text is primarily the product of AI. 
These research efforts entail training detection models~\cite{AITextClassifier,zhan2023g3detector,hao-etal-2025-learning}, employing features for zero-shot detection~\cite{lavergne2008detecting, mitchell2023detectgpt,yang2023dna,hans2024spotting}, or incorporating specific watermarks during content generation~\cite{kirchenbauer2023watermark,Zhao2023ProvableRW,hou2023semstamp}.

While the current body of research predominantly focuses on identifying content substantially generated by AI, thus optimized for binary detection, real-world AI-assisted  generation often involves varying degrees of human contribution.
% In many practical application scenarios, it is not sufficient to merely detect content primarily generated by AI, rather it is crucial to discern the extent of human contribution.
Therefore, distinct from the detection of AI-generated content, our emphasis is on reliably quantifying human contribution within AI-assisted generation from an informational perspective.

\section{Dataset Construction}
To verify the validity and reliability of the proposed measure of human contribution, we construct a dataset of AI-assisted generation data with known varying levels of human contribution. 
% \new{Note that obtaining a dataset with clear, real-valued labels poses a significant challenge, as there is no established method for quantifying human contribution when both human input and output are known.
% }
By design, our dataset spans a very large range of human contributions in AI-assisted output with distinct levels that are hardly controversial. 
For a comprehensive evaluation, we further vary three factors, beyond the level of human contribution:
(1) domains, focusing on those where originality protection is crucial;
(2) different LLMs; and
(3) different random generation runs. 
Building this dataset primarily involves two steps: \emph{raw information collection and processing} and \emph{AI-assisted content generation}.

\textbf{Raw Information Collection and Processing:}
First, we collect and process \emph{multi-level information} in various domains. Specifically, we sample raw data from public datasets across the following domains:  academic paper abstracts, news articles, patent abstracts, and poems. We sample 2,000 entries for each domain. For paper abstracts, each raw data entry includes content, title, and subject; for the other three domains, each raw entry includes content and title. Details of the original dataset and sampling process are provided in section~\ref{supp-raw}.
We further process the data into 
a uniform structure with decreasing levels of information: \emph{content}, \emph{summary}, \emph{title}, and \emph{subject} (except poems, because of less-informative titles), with missing parts of the raw data supplemented using GPT-3.5. The corresponding statistics are presented in Table~\ref{tab:stat}.
% To circumvent the possibility of LLMs being trained on this content andaffecting the evaluation of human contributions, we have gathered the most recent content likely to have been published after the development of the LLMs utilized in our experiments. 
% The specific data collection process, including content sources and the timeframe for data acquisition, is elaborated upon in the Supplementary Materials.

\textbf{AI-Assisted Content Generation:}
Next, we generate new content using LLMs with varying levels of \emph{\textit{human input}} constructed from the earlier process, categorized as follows: \emph{\textbf{polishing}}, \emph{\textbf{generation with summary}}, \emph{\textbf{generation with title}}, \emph{\textbf{generation with subject}} (where applicable). These inputs use information corresponding to content, summary, title, and subject, respectively. The detailed prompt constructions are shown in section~\ref{supp-prompt}. These human inputs represent varying levels of human contribution, from high to low, based on the amount of information provided.
To support a comprehensive analysis, we apply different LLMs, including the state-of-the-art open-weight LLMs Llama-3~\cite{touvron2023llama}
and Mixtral~\cite{mixtral2023}  and the proprietary models GPT-3.5~\cite{link_chatgpt}, GPT-4o, Gemini-2.0 Flash~\cite{team2023gemini}, and  Clause-3.5 Haiku~\cite{claudemodelcard}. The specific API versions of the models are detailed in section~\ref{supp-modelsetup}. We generate five times for a human input with the temperature set as $0.7$ for diverse outputs. 

\section{Methods}

% \subsection{Defining Human Contribution in AI-Assisted Generation }
% In contrast to a binary classifier determining whether content is primarily generated by AI, our aim is to derive a quantitative measurement indicating the extent of human contribution in AI-assisted content generation. 
% This metric enables us to more effectively apply it in the refinement of copyright law and policy regarding AI-assisted generation. 
% This is a novel and previously unexplored issue. 
\subsection{Defining Human Contribution in AI-Assisted Generation}
Our core idea revolves around utilizing \textit{information content} as a medium for gauging the contributions of humans and AI. Particularly, we define human contribution in AI-assisted generation as the ratio of mutual information between human input and AI-assisted output relative to the total self-information of the AI-assisted output, as illustrated in Figure~\ref{fig:overview}\textbf{b}.

In this section, we first introduce related concepts derived from information theory~\cite{shannon1948mathematical}; we then provide our definition of human contribution.
In the following definition, we consider an AI model $M_\theta$, its generative distribution $p_{\theta}$, human input $\boldsymbol{x}$, and AI-assisted output $\boldsymbol{y}$.

First, we quantify the information content within the generated output $\boldsymbol{y}$ through the concept of \emph{self-information}. Self-information measures the level of surprisal associated with an event, reflecting the probability of that event occurring. In this context, generating content that is less probable is considered more informative event. We represent the self-information of the generated output $\boldsymbol{y}$ as follows:
\begin{equation}
    I(\boldsymbol{y}) = -log(p_{\theta}(\boldsymbol{y})),
\end{equation} 
where $p_{\theta}(\boldsymbol{y})$ is the probability that the content $\boldsymbol{y}$ is generated without any condition.

On the other hand, when conditioned on human input  $\boldsymbol{x}$, the information content within the generated output $\boldsymbol{y}$ transforms into \emph{conditional self-information}. Conditional self-information quantifies the information contained in an event, given the occurrence of another event. Here, we represent the conditional self-information of the generated output $\boldsymbol{y}$ given ``human input is $\boldsymbol{x}$'' as follows:
\begin{equation}
    I(\boldsymbol{y}\mid\boldsymbol{x}) = -log(p_{\theta}(\boldsymbol{y}\mid\boldsymbol{x})                                    ),
\end{equation}
where $p_{\theta}(\boldsymbol{y}\mid\boldsymbol{x})$ is the probability that the content $\boldsymbol{y}$ is generated conditioned on human input $\boldsymbol{x}$. 
% It is noteworthy that both information and conditional information are calculated for the specific event “generated output is $\boldsymbol{y}$,” rather than the entire generative distribution—which would be difficult to measure due to the vast space of all possible generated contents. This allows the probabilities  $p_{\theta}(\boldsymbol{y})$ and $p_{\theta}(\boldsymbol{y}\mid\boldsymbol{x})$ to be easily computed using the corresponding generative probabilities, without and with human input. 
Specifically, considering the generative process of the autoregressive LLM $M_\theta$, these probabilities can be computed by multiplying the probabilities of each token $y_i$ being sampled from the generative distribution, conditioned on the previously generated tokens (and the human input $\boldsymbol{x}$ for conditional probability with human input):
\begin{align}
&p_\theta(\boldsymbol{y}) = \prod_{i=1}^{N} p_\theta(y_i \mid \boldsymbol{y}_{<i}), \\
&p_\theta(\boldsymbol{y} \mid \boldsymbol{x}) = \prod_{i=1}^{N} p_\theta(y_i \mid \boldsymbol{y}_{<i}, \boldsymbol{x}),
\end{align}
\new{where $N$ is the total number of tokens in $\boldsymbol{y}$}. 
% \re{Note that all probabilities are computed at the token level.}

\new{Based on these two information concepts, we define \emph{mutual information} between the generated content $\boldsymbol{y}$ and the human input $\boldsymbol{x}$ as
the reduction in surprisal when human input $\boldsymbol{x}$ is known for generating content $\boldsymbol{y}$. This indicates how much information content in $\boldsymbol{y}$ can be attributed to human input  $\boldsymbol{x}$. Formally, it is defined as:}
% Then we consider the \emph{mutual information} between the generated content $\boldsymbol{y}$ and the human input $\boldsymbol{x}$. Mutual information measures the reduction in uncertainty about one variable given knowledge of another, reflecting the amount of dependence between the two variables. Here, the mutual information indicates the reduction of surprise (information) when human input $\boldsymbol{x}$ is given in generating the content $\boldsymbol{y}$, defined as:
\begin{equation}
    I (\boldsymbol{x}; \boldsymbol{y}) = I (\boldsymbol{y}) - I (\boldsymbol{y}\mid\boldsymbol{x}).
\end{equation}

Building upon the aforementioned definition of information within the AI-assisted generation process, we proceed to establishing the definition of human contribution in AI-assisted generation. 

% From information theory prospective, we would like to use the generated content $\boldsymbol{y}$'s information reduction because of human input $\boldsymbol{x})$ as the human contribution, and 

% For each token, the $t$-th token is sampled from the \emph{AI-assisted distribution} $p_{\theta}(\cdot | \boldsymbol{y}_{<t}, \boldsymbol{x})$, which is
% conditioned on both the human input $\boldsymbol{x}$ and the sequence of previously generated tokens $\boldsymbol{y}_{<t}$.
% In contrast, when the human input $\boldsymbol{x}$ is unknown, the \emph{model 
% distribution} $p_{\theta}(\cdot | \boldsymbol{y}_{<t})$  represents the distribution without the knowledge of human input $\boldsymbol{x}$.

% From an information theory perspective, we propose that the reduction in entropy during the generation of output $\boldsymbol{y}$ conditioned on human input $\boldsymbol{x}$ relative to when $\boldsymbol{x}$ is unknown serves as a measure of the mutual information between the human input and the AI-assisted generated content. 
% This reduction in entropy, when normalized by the total entropy when $\boldsymbol{x}$ is unknown, quantifies the relative mutual information between the human input and the AI-assisted content. 
% We define this ratio as the \emph{human's contribution $\phi$}.
\begin{figure*}[t]
    \centering
    \includegraphics[width=1.0\linewidth]{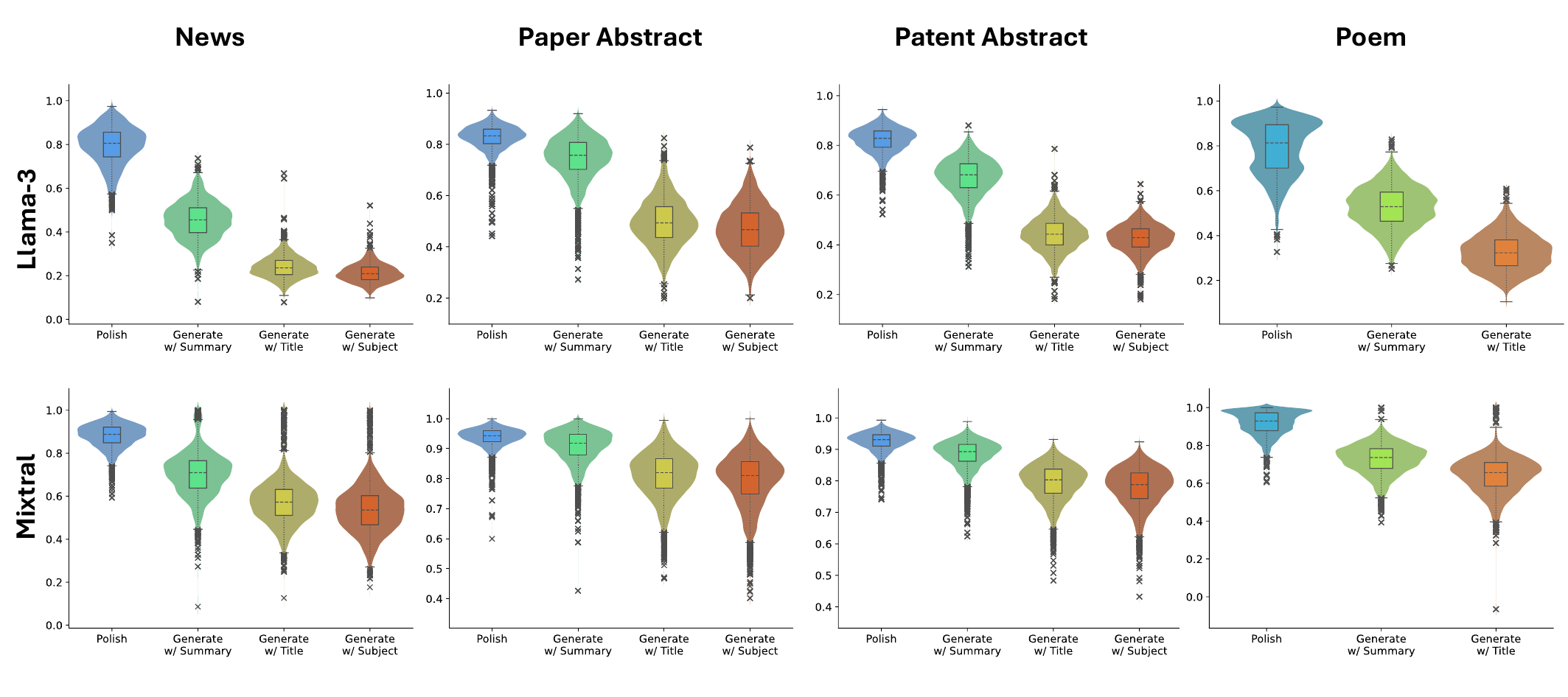}
    \caption{
 The distribution of the outcomes of the \textbf{proposed measure} for the constructed dataset.  Overall, the proposed measure exhibits the expected trend that lower values are obtained for the generated content with less human contribution.
} 
    \label{fig:self}
\end{figure*}

\begin{definition}[Human contribution]
Given an AI model $M_\theta$ and human input $\boldsymbol{x}$, where $\boldsymbol{y}$ represents the AI-assisted generated content, the human contribution $\phi$ is defined as the ratio of mutual information $I (\boldsymbol{x}, \boldsymbol{y})$ to self-information $I(\boldsymbol{y})$:
$\phi = \frac{I(\boldsymbol{x}; \boldsymbol{y})}{I(\boldsymbol{y})}.$

\end{definition} 

This definition of human contribution pertains to the proportion of the information content within the generated output that can be attributed to human input, relative to the total information content of the generated output.
% \new{This definition can be directly employed for evaluation in scenarios where human input is available and reliable.
% For instance, model service providers can apply this definition during the generation process to quantify human contributions and issue certified documentation.
% Moreover, in evidentiary scenarios for originality authentication, human authors can provide reliable human input (certified with appropriate documentation) for evaluation. }
This definition can be directly employed for \emph{originality authentication}, where the human author can provide both the AI model's generative distribution and the original human input (certified with appropriate documentation) for authentication and evaluation, as well as cases where \emph{model service providers} directly apply the definition during the generation process to calculate a measurement of human contribution.

\subsection{Estimating Human Contribution Without Human Input }
\label{sec:estimation}
To better address real-world scenarios where the AI-assisted generation process is unknown, we extend our method to estimate the human contribution of any given text alone without knowledge of human input and output generation processes.
We have less information now than the situation considered before. This loss of information leads to a degree of uncertainty, as the same output can be generated from inputs with varying levels of contribution.
Therefore, we redefine our goal 
to estimate the proportion of information content in the AI-assisted output that is \textit{necessarily} provided by humans.  This is also referred to as the \textit{minimum human contribution}  for human input  $\boldsymbol{x}$  to plausibly generate the AI-assisted output  $\boldsymbol{y}$.  
This focus on minimal human contribution stems from several practical considerations, such as evaluating the unique original informativeness value that LLMs cannot easily generate, as detailed in section~\ref{supp-method}.
% \re{We acknowledge that, under this reframed research objective, the measured value no longer represents an exact value of human contribution, but rather a meaningful lower bound.}
% }

To compute this, we adapt the original formula  $\frac{I(\boldsymbol{x}; \boldsymbol{y})}{I(\boldsymbol{y})} $. While the denominator  $I(\boldsymbol{y})$  can still be directly calculated, the numerator  $I(\boldsymbol{x}; \boldsymbol{y}) = I(\boldsymbol{y}) - I(\boldsymbol{y}|\boldsymbol{x})$  is estimated as its minimum possible value for a valid hypothetical $ \boldsymbol{x}$  that plausibly generates  $\boldsymbol{y} $.
Intuitively, if a hypothetical  $\boldsymbol{x}$  plausibly generates  $\boldsymbol{y} $, the likelihood of  $\boldsymbol{y} $ being generated based on $ \boldsymbol{x} $, $p_\theta(\boldsymbol{y} \mid \boldsymbol{x})$, should be sufficiently high (have a lower bound). This implies that the conditional information (or surprisal)  $I(\boldsymbol{y}|\boldsymbol{x}) = -log(p_{\theta}(\boldsymbol{y}\mid\boldsymbol{x})                                    )$ of generating $ \boldsymbol{y}$  given  $\boldsymbol{x}$  should have an upper bound. We leverage this upper bound to compute the minimum possible value of  $I(\boldsymbol{x}; \boldsymbol{y})$.
Due to limited space, the detailed motivation for estimating minimum human contribution, along with the calculation method, is provided in section~\ref{supp-method}.
\section{Experiments}

\subsection{Human Contribution Evaluation}
\label{sec:exp-main}
We evaluate the effectiveness of the proposed measure using the constructed dataset. 
This section focuses on the original scenario where both the AI-assisted output $\boldsymbol{y}$ and human input $\boldsymbol{x}$ are known, and the AI model $M_\theta$'s output probability is available in evaluation. 

\textbf{\new{Overall Trend Analysis:} }Figure~\ref{fig:self} illustrates the human contribution results of two open-weight LLMs, Llama-3 and Mixtral, on the constructed dataset across various domains. 
From the results, we make the following observations.
First, for each combination of model and data domain, varying levels of human contribution yield different distributions for measured human input in the expected direction: the lower the human author's informational contribution in AI-assisted generation (from polishing, to generation with summary, to generation with title, and finally to generation with subject), the smaller the proposed metric's value.
% For instance, when generating news articles with Llama-3, polished content demonstrates an average human contribution measurement of $85.37\%$, while content generated with a subject shows an average human contribution measurement of $30.83\%$. 
This indicates that our proposed measure can effectively distinguish different levels of human contribution in AI-assisted generation, providing useful measurements from an informational perspective.

\begin{figure*}[t!]
    \centering
    \includegraphics[width=1.0\linewidth]{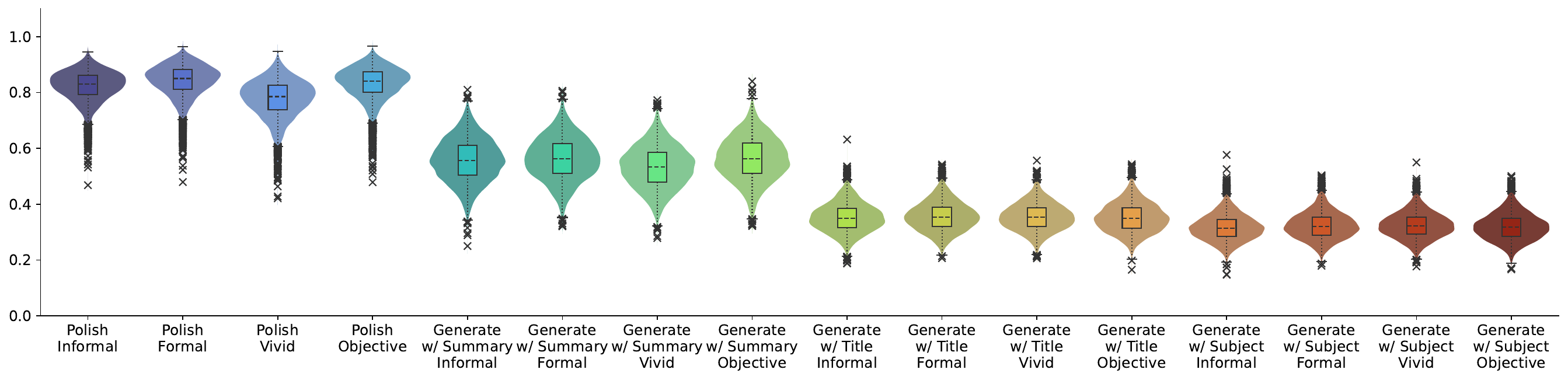}
    \caption{
 \re{The distribution of the outcomes of the proposed measure for the constructed dataset of news in different writing styles using Llama-3. Overall, the writing styles have little influence on the measurement outcomes.}
} 
    \label{fig:style}
\end{figure*}
\begin{figure*}[t!]
    \centering
    \includegraphics[width=1.0\linewidth]{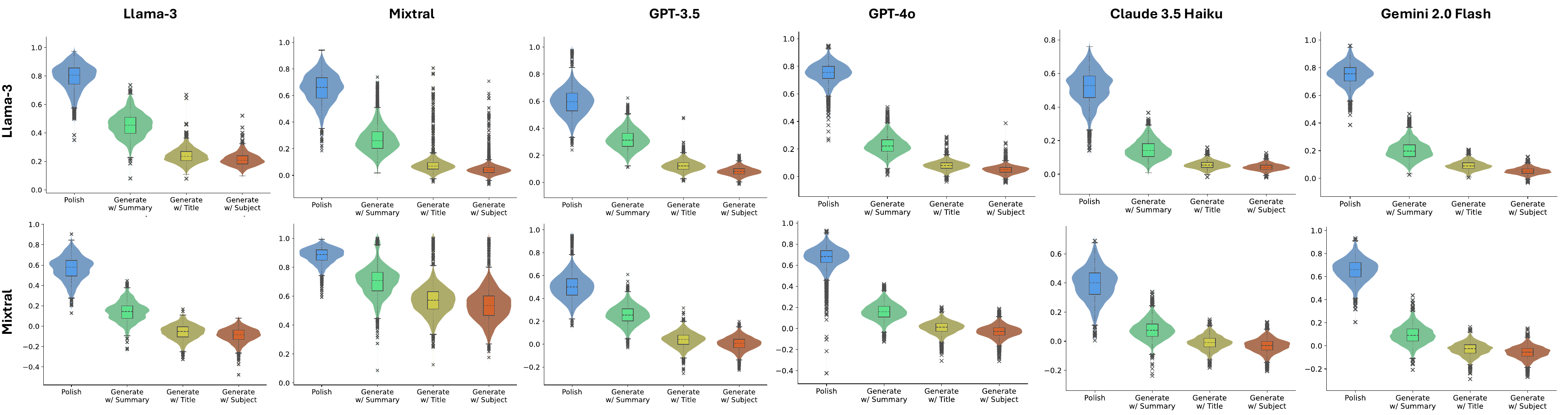}
    \caption{
 The distribution of the outcomes of the proposed measure for the constructed dataset of news articles for various generation models (columns) and surrogate models for measurement (rows). Overall, for each model pair, the proposed measure exhibits the expected pattern that lower measured values are obtained for the generated content with less human contribution.
} 
    \label{fig:other-news}
\end{figure*}

Second, we observe variability in the outcomes of the proposed measure across different data domains for a specific generation mode (e.g., generation with summary). These differences are reasonable because the same generation mode does not necessarily equate to a similar percentage of human contribution across different domains. 
\new{For example, in our dataset, the News category consists of full articles with detailed and vivid descriptions, while Paper Abstracts are typically concise paragraph. However, the summaries, titles, and subjects used for generation in our dataset for both categories are of comparable information content. Consequently, when generating a news article versus a paper abstract from a summary, title, or subject, the expected human contribution distributions differ due to the inherent difference in the information content of the data.
}
% As discussed in the Dataset Construction section, we can only generate content with approximate varying levels of human contribution.
% For example, the ratio of summary length to content length is significantly higher for paper abstracts than for news articles, as shown in Table~\ref{tab:stat}, indicating a variation in ground truth human contribution. 
Therefore, in our evaluation, the primary consideration is to verify whether our measure can consistently reflect the overall pattern of varying levels of human contribution for each generation model and creative domain. This consistency would validate the reliability of our proposed measure for distinguishing different levels of human contribution.

\new{\textbf{Human-annotation Dataset:} To further validate the reliability and consistency of our quantitative analysis, we construct a dataset with human annotations. 
As discussed earlier, obtaining a dataset with clear, real-valued labels remains a significant challenge, as there is no established method for quantifying human contribution when both human input and output are known. Inspired by evaluation platforms such as Chatbot Arena~\cite{chiang2024chatbot}, we collect human annotation data by pairwise comparisons.
Specifically, we sample \re{$1,500$} pairs of input-output data points generated by Llama-3 across all four domains and categories with the condition that the measured human contribution gap between the two points exceeds 0.1 using our method. This threshold is chosen to ensure the meaningfulness of the comparisons and to reduce ambiguity during annotation. Pairs of input-output data points are so annotated \re{by three human annotators} as to reflect which one contains a higher degree of human contribution. \re{The annotation process is detailed in section~\ref{supp-human}}.
Using this annotated dataset, we evaluate how well the relative magnitudes derived from our measurements align with human judgments in these comparisons. The results indicate that \textbf{95.93\%} of the data exhibit consistency between our measurements and human annotations. Moreover, as shown in Figure~\ref{fig:gap}, the inconsistent pairs, \re{i.e., cases where the human annotation disagrees with the measurement’s ranking,} are predominantly concentrated in cases where the measured human contribution gap between two data points is less than 0.2.
This further lends support to the robustness and validity of our measurement, demonstrating its consistency across different domains and setups.
}

Due to space limitations, we defer the analysis of \textbf{the impact of content length} (\ref{supp-length}), \textbf{model temperature} (\ref{supp-temp}), and \textbf{resilience to adaptive attacks} (\ref{supp-adaptive}) to the Appendix.
% Consequently, it is difficult for human evaluators to directly assess the correctness of our measurement results for individual input-output data points.

\subsection{Impact of Writing Style}
\label{supp-style}
\re{Note that our proposed framework is based on token-level probability calculations. As a result, semantically similar content may yield different measured values of human contribution due to variations in writing style.
In this section, we conduct an ablation study to examine the impact of writing style on the outcomes of our measurement using Llama-3.  Specifically, we explore four stylistic variations within the news domain by appending the following prompts:
“Generate in an informal style.”, “Generate in a formal style.”, “Generate in a vivid story-telling style.”, and “Generate in an objective news reporting style.” 
These prompts are applied while keeping the original input content fixed, thereby inducing markedly different wording styles without altering other human contributed information. This setup allows us to assess the extent to which semantically-preserving stylistic variation influences the outcomes of our measurement.}

\re{The results are shown in Figure~\ref{fig:style}. The results demonstrate that the proposed measurement robustly captures differences in human informational contribution, even when writing style varies. This is consistent with our expectation, as the variance introduced by such semantically-preserving stylistic fluctuations is relatively limited compared to the differences driven by variations in the amount of information provided in the input.}

\subsection{Generalization of Our Method}

In real-world applications, the AI model's generative probability  
 $p_{\boldsymbol{\theta}}$
  may not be available. For instance, applications like ChatGPT don't release generative probabilities to users.
This section demonstrates whether a surrogate model with generative probability 
$p_{\boldsymbol{\theta}}^\prime$ 
  can be employed for our assessment when the AI model's generative probability 
 $p_{\boldsymbol{\theta}}$
  is unknown. 
Specifically, in this experiment, we use Llama-3 and Mistral as the surrogate models and use their generative probability 
$p_{\boldsymbol{\theta}}^\prime$ to assess the content generated by various LLMs (Llama-3, Mistral, \re{GPT-3.5, GPT4o, Claude 3.5 Haiku, and Gemini 2.0 Flash}) in the constructed dataset. \new{Note that generation models can include chatbots or models that do not release their generative probabilities, while surrogate models require access to generative probability distributions for evaluation.}
% \re{Note that in this setting, due to the lack of access to the true generative distribution, the evaluated result is no longer an exact human contribution value as defined in the generative distribution-known setting. Instead, similar to the estimation framework, it serves as a meaningful quantitative reference.}

\begin{figure*}[t!]
    \centering
    \includegraphics[width=0.96\linewidth]{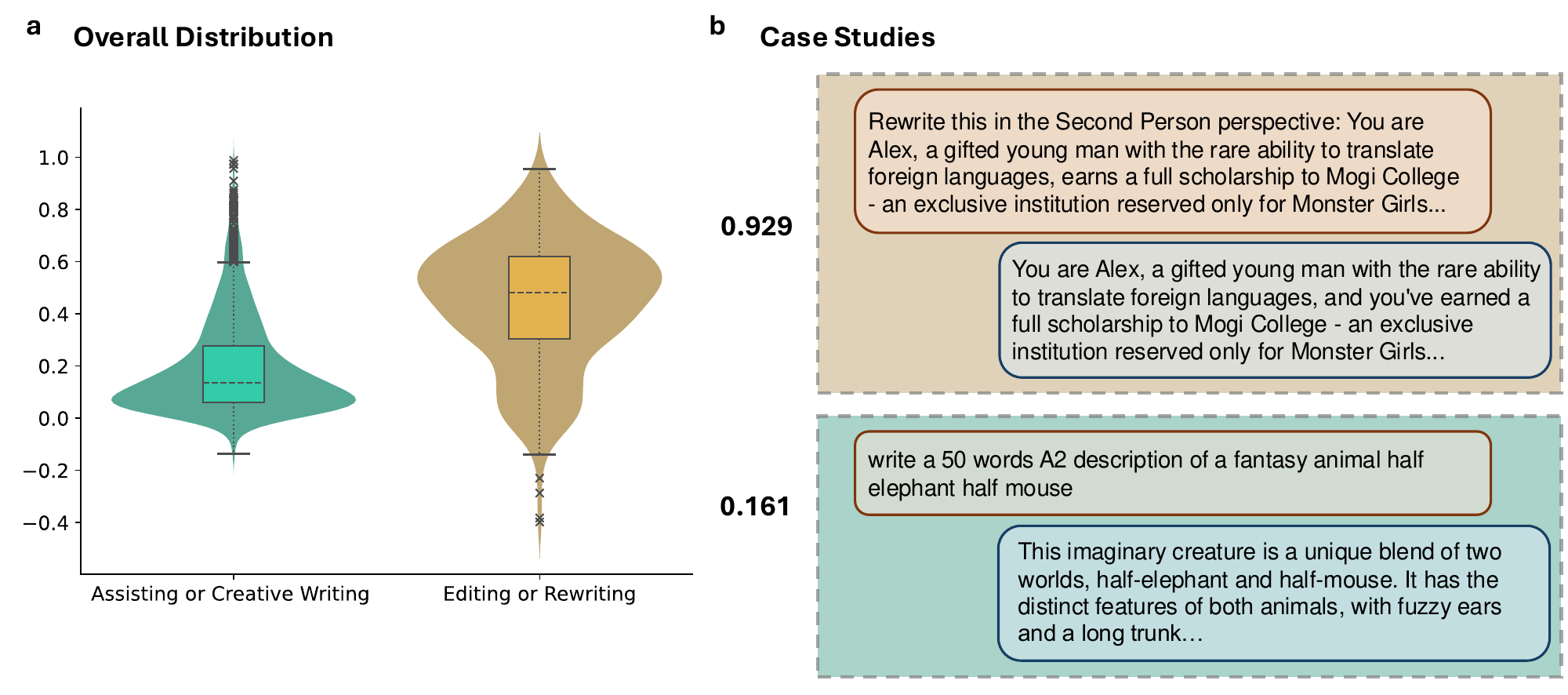}
    \caption{
 \textbf{a.} The distribution of the outcomes of the proposed measure across various classes of real-world AI-assisted generation, sampled from WildChat.  The ``editing or rewriting" class typically exhibits a higher degree of measured human contribution.
 \textbf{b.} Case studies on data samples from two distinct classes in WildChat.
} 
    \label{fig:application}
\end{figure*}
% \re{\textbf{Overall Trend Analysis:}} 
Figure~\ref{fig:other-news} illustrates the effectiveness of our approach in the news domain for various combinations of surrogate model (rows) and generation model (columns). 
Results for other domains are presented in section~\ref{supp-generalization}.
We observe that even without using the original AI model for evaluation, our proposed measure captures the expected trend in human contribution across various surrogate and generative model combinations. This validates the applicability of our measure when generation model information is unavailable.
This effectiveness may be attributed to the similarity in generative distributions across LLMs, stemming from the universal knowledge they share during training. The gradient in human contribution across varying levels of human input is far more pronounced than the differences between the distributions themselves, indicating that our method is a robust assessment tool.

\subsection{Applications to Real-World AI-Assisted Generation }
The aforementioned experiments were conducted on a synthetic dataset with known varying levels of human contribution, allowing us to verify the reliability of our measurement method. To test its real-world applicability, we apply our method to real-world scenarios involving user interactions with LLMs.
Specifically, we sample cases from the WildChat-1M dataset~\cite{zhao2024wildchat} and classify them using a prompt classification tool~\cite{huggingface2024}. We then sample data from two prompt classes related to AI-assisted generation: ``assisting or creative writing'' (2,000 entries) and ``editing or rewriting'' (500 entries), according to their counts in the dataset. The evaluation surrogate model is Llama-3, while the contents were generated with ChatGPT.

Figure~\ref{fig:application}\textbf{a} demonstrates the overall distribution of measured human contributions across the two classes. We expect that the ``editing or rewriting'' class will involve more human contribution than ``assisting or creative writing.'' Consistent with this, the measured human contributions are generally higher for ``editing or rewriting.'' We present two specific cases for two classes respectively in Figure~\ref{fig:application}\textbf{b}. Overall, the measured human contributions align with our expectations. For instance, the ``editing or rewriting'' case is measured as having 92.86\% human contribution, while the ``assisting or creative writing" case is measured at 16.14\%. These distribution and case study results further support the validity of our method in measuring human contribution in real-world AI-assisted generation contexts.
\begin{figure*}[t!]
    \centering
    \includegraphics[width=1.0\linewidth]{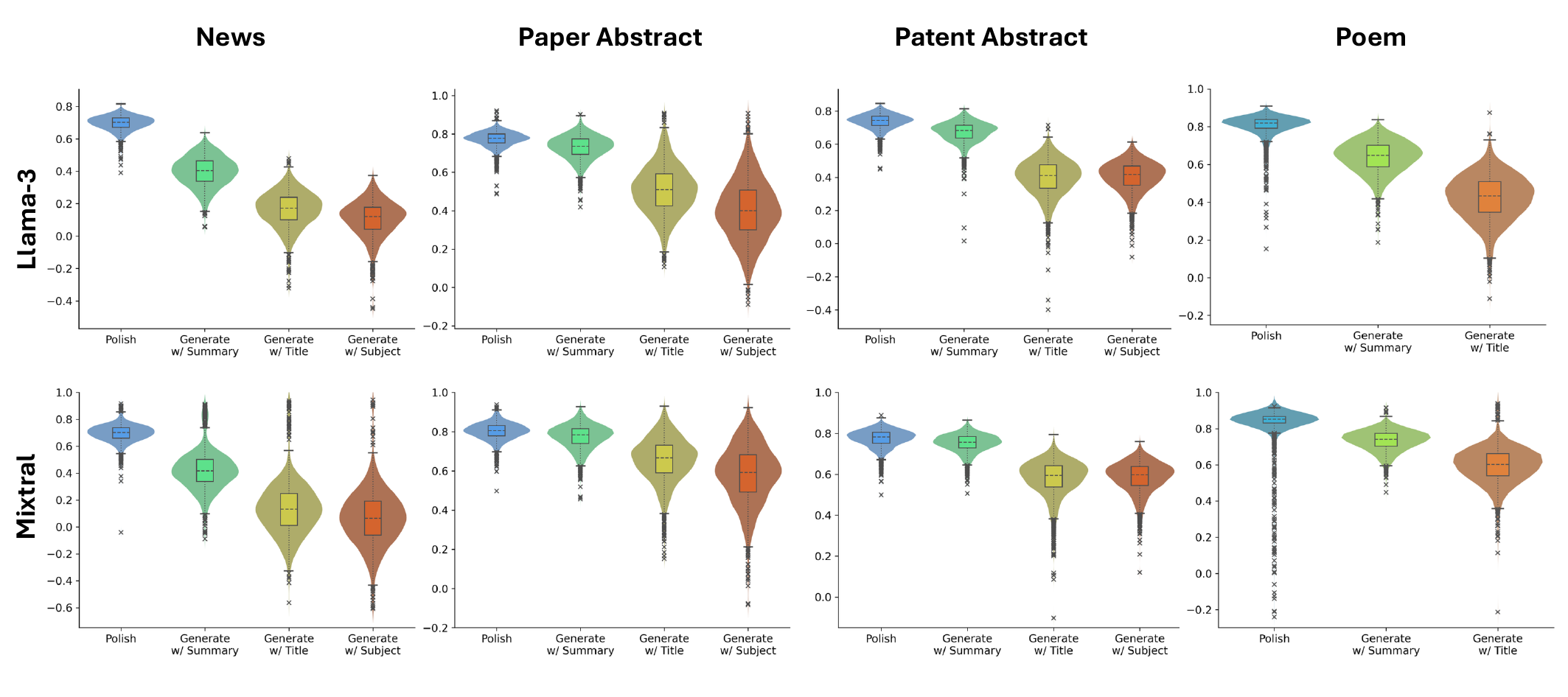}
    \caption{
 \new{The distribution of the outcomes of the \textbf{proposed estimation} for the constructed dataset.  Overall, the proposed estimation exhibits the expected trend that lower values are obtained for the generated content with less human contribution.}
} 
    \label{fig:estimation}
\end{figure*}

\subsection{\new{Human Contribution Estimation Without Human Input}}

% % \begin{table}[t]
% %     \centering
% %     \begin{tabular}{c|c|cccc|c}
% %     \toprule
% %           &  & News & Paper Abstract & Patent Abstract & Poem & Merged \\
% %          \midrule
% %          \multirow{2}{*}{MSE} 
% %          & Llama-3  & 0.0389 & 0.0216 &  0.0248& 0.0079 & 0.0270  \\
% %          & Mixtral  &0.1606  &0.0330  & 0.0373 & 0.0209 &  0.0715\\
% %          \midrule
% %          \multirow{2}{*}{Spearman's rho} 
% %          & Llama-3  & 0.8781 & 0.8937 & 0.8376 &0.9314  & 0.9026 \\
% %          & Mixtral  & 0.7463 &0.8303  & 0.7337 & 0.8031 & 0.8346 \\
% %          % \midrule
% %          % \multirow{2}{*}{AUC} 
% %          % & Llama-3  & 0.8610 &0.8659 &  0.8268  & 0.8907 & 0.7458 \\
% %          % & Mixtral  &0.7900  &0.8289&  0.7794&   0.8197 & 0.6637 \\
% %     \bottomrule
% %     \end{tabular}
% %     \caption{Comparison between estimation without human input and evaluation with human input.}
% %     \label{tab:my_label}
% % \end{table}

% \new{\textbf{Overall Trend Analysis:}
{
In Figure~\ref{fig:estimation}, we present the human contribution estimation results for two state-of-the-art open-source LLMs, Llama-3 and Mixtral, applied to the constructed dataset across various domains. We make the following observations:
First, the proposed estimation method produces results that follow the expected trend in our dataset: as the human author’s informational contribution decreases during AI-assisted generation, the estimated metric value correspondingly decreases.
Second, as we suppress the diverse real values of $I(\boldsymbol{y} \mid \boldsymbol{x})$ to an estimated boundary value that is empirically chosen, the resulting distributions do not fully match human contribution evaluations where $\boldsymbol{x}$ is known (Figure~\ref{fig:self}).
Finally, since this is an estimation process of a lower bound, there are cases where the estimated values fall below zero. In our original framework, under the defined setting where both the human input and the generative distribution are known, the value range is generally between 0 and 1.
However, due to the nature of lower-bound estimation, some values may fall below zero. These negative estimates can be interpreted as cases where the corresponding outputs could plausibly have been generated with minimal or negligible human contribution, under the minimal contribution assumption. 
% We further explore case studies in Supplementary Material section 2.7.

In addition, we use the annotated dataset same as section~\ref{sec:exp-main} to evaluate how well the relative magnitudes derived from our estimates align with human judgments in these comparisons. The results indicate that \textbf{92.87\%} of the samples exhibit consistency between our estimates and human annotations, further supporting the validity of our estimates.

We further demonstrate the effectiveness of the proposed estimation method in \textbf{multi-round generation scenarios} in section~\ref{supp-multi} and \textbf{AI-generated content detection} in section~\ref{supp:detection}.

Overall, our method provides a meaningful solution for estimating human contribution in situations where direct input information and generation processes are unavailable.}

\section{Conclusion}

This study formally frames the challenge of quantifying human contribution in AI-assisted generation and introduces a principled, information-theoretic method as a credible measure. We validate the proposed measure through extensive experiments on multi-domain AI-assisted generation datasets using multiple LLMs. To enhance real-world applicability, we further extend the framework to scenarios in which the original human input is unavailable and or the underlying generative distribution is inaccessible.
By articulating this research question and proposing a principled measurement framework, this work aims to advance future efforts in originality delineation and the regulation of AI-assisted content in generative AI era.

\section*{Limitations}
While it serves to frame the question and provide a preliminary method for measuring human contribution, our work has several limitations. 
% First, we currently focus on scenarios where human input is available and reliable for evaluation. Further research is necessary to measure human contributions in scenarios where human input is unknown.
% This is applicable in contexts where LLM service providers can specify the details of human contribution or where individuals are willing to report their inputs. However, there is currently no legal requirement for copyright owners or registration applicants to declare their authorial methods. In its current form copyright registration  relies  only on the content of the registered work even if that work's generation was AI-assisted. This could be addressed with a combination of legal and technical advancements. On the legal side—bracketing complexities that arise from multilateral agreements that constrain alterations to copyright law—copyright administrators could mandate a declaration of prompt details for AI-assisted works. On the technical side, researchers could explore whether input derived from prompt inversion can be effectively applied within this framework, as we hope to do in future work.
% Second, our current validations are primarily conducted on synthetic datasets, which present a limited range of scenarios. 
% We aim to apply our measure in more real-world contexts to further optimize and understand its performance.
First, the current framework focuses on textual output from LLMs. However, originality issues related to AI-assisted generation are not limited to text; they also extend to images, audio, video, computer code, etc. Incorporating non-textual output raises even more complex problems due to the change in modality between human input and AI-assisted output. We aim to explore originality issues in such scenarios in future research.
\new{In other modalities, since data may no longer be discrete, this framework may not be directly applicable. However, as generative models are fundamentally grounded in certain generative probabilistic distributions, we hope that the information theory-based framework proposed in this work can serve as a foundation and inspiration for addressing originality challenges across  diverse modalities.} 
Second, caution is warranted when measuring human contribution in AI-assisted content generation in the copyright domain. Human edition, selection, and compilation of AI-assisted content may provide significant creative input, which could be relevant when assessing authorship for copyright purposes.
% \re{Third, some of the original source data used to construct our dataset may have been seen by the LLMs during pretraining, potentially influencing the experimental results. To further validate the robustness of our method, we collected a new set of 1,000 news articles produced in 2024—after the training cutoff of the models used in our experiments—and repeated the same generation and evaluation process. The results, presented in section 2.2 in Supplementary Materials, exhibit trends consistent with those observed in the main experiments, further supporting the validity of our approach.}
\new{Finally, we acknowledge that “human contribution” can have different meanings. In this work, we focus specifically on measuring informational contribution, while leaving other interpretations, such as prompt-engineering effectiveness, for future exploration. 
Also, informational contribution should not be directly equated with creativity or originality. Our framework is designed to quantify the informational contribution contained in the input, rather than to provide a normative judgment about creative value or authorship.
}

\section*{Ethical Considerations}
The objective of this study is to pose a research question and propose a framework for measuring human contribution in AI-assisted content generation. 
This question and framework aim to facilitate originality delineation in the era of creation with the assistance of AI. 
Simultaneously, this work seeks to inspire more research on technical methods that can support the enhancement of relevant regulations in the context of widespread AI utilization in various contexts.
A potential risk is that in real-world applications of the proposed framework, there might be targeted adaptive attacks aimed at manipulating the results to artificially elevate the assessed level of human contribution. 
Although this paper examines two adaptive attacks and verifies the robustness of the proposed measure against them, more sophisticated and advanced attacks may arise in real-world scenarios.
We hope to further understand and mitigate such risks in future work.

The authors of this paper introduce a  method to technically measure  human contribution in AI-assist content generation. The method is widely applicable in different situations. However, the paper does not intend to discuss the complex copyright legal and policy issues related to “originality” or “eligibility,” nor it reflect any of Microsoft’s legal and policy positions on the copyright issues. 
% \section*{Acknowledgments}

% This document has been adapted
% by Steven Bethard, Ryan Cotterell and Rui Yan
% from the instructions for earlier ACL and NAACL proceedings, including those for
% ACL 2019 by Douwe Kiela and Ivan Vuli\'{c},
% NAACL 2019 by Stephanie Lukin and Alla Roskovskaya,
% ACL 2018 by Shay Cohen, Kevin Gimpel, and Wei Lu,
% NAACL 2018 by Margaret Mitchell and Stephanie Lukin,
% Bib\TeX{} suggestions for (NA)ACL 2017/2018 from Jason Eisner,
% ACL 2017 by Dan Gildea and Min-Yen Kan,
% NAACL 2017 by Margaret Mitchell,
% ACL 2012 by Maggie Li and Michael White,
% ACL 2010 by Jing-Shin Chang and Philipp Koehn,
% ACL 2008 by Johanna D. Moore, Simone Teufel, James Allan, and Sadaoki Furui,
% ACL 2005 by Hwee Tou Ng and Kemal Oflazer,
% ACL 2002 by Eugene Charniak and Dekang Lin,
% and earlier ACL and EACL formats written by several people, including
% John Chen, Henry S. Thompson and Donald Walker.
% Additional elements were taken from the formatting instructions of the \emph{International Joint Conference on Artificial Intelligence} and the \emph{Conference on Computer Vision and Pattern Recognition}.

\bibliography{custom}

\appendix
\clearpage
% \renewcommand{\contentsname}{Appendix Contents}
% \setcounter{tocdepth}{2}
% \newpage
\tableofcontents
% \section{Appendix}
\label{sec:appendix}
\clearpage
% \tableofcontents
% \subsection{Multi-round Human }
\section{Additional Experimental Setups}
\subsection{Raw Data Collection and Processing}
\label{supp-raw}
\subsubsection{Raw Data Collection}
This section details the collection and processing of the raw data.
\begin{itemize}
    \item \textbf{Paper Abstract:} The paper abstract data entries are randomly sampled from the \textit{arxiv-abstracts-2021} dataset\footnote{\url{https://huggingface.co/datasets/gfissore/arxiv-abstracts-2021}}, containing the original paper abstracts as content, titles as title, and primary categories as subject. The summaries are generated using GPT-3.5 using the prompt specified in Section~\ref{supp:data_process}.
    \item \textbf{News:} The news data entries are sampled from the \textit{News Articles} dataset\footnote{\url{https://dataverse.harvard.edu/dataset.xhtml?persistentId=doi:10.7910/DVN/GMFCTR}}, containing the original news articles as content and titles as title. To avoid outliers, the entries are randomly sampled with a content length constraint of 400 to 700 words. The summaries and subjects are generated using GPT-3.5 with the prompt specified in Section~\ref{supp:data_process}.
    \item \textbf{Patent Abstract:} The patent abstract data entries are sampled from the \textit{HUPD 2018} dataset\footnote{\url{https://huggingface.co/datasets/HUPD/hupd/blob/main/data/2018.tar.gz}}, containing the original patent abstracts as content and titles as title. To avoid outliers, the entries are randomly sampled with a content length constraint of 150 to 250 words. The summaries and subjects are generated using GPT-3.5 with the prompt specified in Section~\ref{supp:data_process}.
    \item \textbf{Poem:} The poem data entries are sampled from the \textit{Poetry Foundation} dataset\footnote{\url{https://www.kaggle.com/datasets/tgdivy/poetry-foundation-poems}}, containing the original poems as content and titles as titles. To avoid outliers, the entries are randomly sampled with a content length constraint of 150 to 250 words. The summaries are generated using GPT-3.5 with the prompt specified in Section~\ref{supp:data_process}. Since the titles already contain a small amount of information, averaging 3.65 words, we does not generate additional subjects.
\end{itemize}
% The prompts for summary 
\subsubsection{Raw Data Processing}
\label{supp:data_process}
For the missing information, including summaries and subjects, we used GPT-3.5 (\textit{gpt-3.5-turbo-1106}) to supplement them, thereby facilitating AI-assisted generation with multiple levels of human contribution. The prompts are specified as follows:
\begin{itemize}
    \item \textbf{Summary Generation:} You are a helpful assistant. Help me summarize the following content in a few sentences as concise as possible: \textit{\{Content\}}.
        \item \textbf{Subject Generation:} You are a helpful assistant. Help me generate the subject of the following content in two to four words: \textit{\{Content\}}.
\end{itemize}
\subsection{Prompt Construction}
\label{supp-prompt}
We detail the prompt construct for AI-assisted generation with different levels of human contribution as follows, where \textit{N} is the original length of the content in the dataset:
\subsubsection{News}
\begin{itemize}
    \item \textbf{Polish:} Help me polish the following news article: \{\textit{News Content}\}. Limit your response to \{\textit{N}\} words. Start with ``News:".
     \item \textbf{Generate with Summary}: Generate a news article with the following summary: \{\textit{News Summary}\}. Limit your response to \{\textit{N}\} words. Start with ``News:".
    \item \textbf{Generate with Title}: Generate a news article with the title: \{\textit{News Title}\}. Limit your response to \{\textit{N}\} words. Start with ``News:".
    \item \textbf{Generate with Subject}: Generate a news article with the subject: \{\textit{News Subject}\}. Limit your response to \{\textit{N}\} words. Start with ``News:".
\end{itemize}
\subsubsection{Paper Abstract}
\begin{itemize}
    \item \textbf{Polish:} Help me polish the following paper abstract: \{\textit{Paper Abstract}\}. Limit your response to \{\textit{N}\} words. Start with ``Abstract:".
     \item \textbf{Generate with Summary}: Generate a paper abstract with the following summary: \{\textit{Paper Summary}\}. Limit your response to \{\textit{N}\} words. Start with ``Abstract:".
    \item \textbf{Generate with Title}: Generate a paper abstract with the title: \{\textit{Paper Title}\}. Limit your response to \{\textit{N}\} words. Start with ``Abstract:".
    \item \textbf{Generate with Subject}: Generate a paper abstract with the subject: \{\textit{Paper Subject}\}. Limit your response to \{\textit{N}\} words. Start with ``Abstract:".
\end{itemize}

\subsubsection{Patent Abstract}
\begin{itemize}
    \item \textbf{Polish:} Help me polish the following patent abstract: \{\textit{Patent Abstract}\}. Limit your response to \{\textit{N}\} words. Start with ``Abstract:".
     \item \textbf{Generate with Summary}: Generate a patent abstract with the following summary: \{\textit{Patent Summary}\}. Limit your response to \{\textit{N}\} words. Start with ``Abstract:".
    \item \textbf{Generate with Title}: Generate a patent abstract with the title: \{\textit{Patent Title}\}. Limit your response to \{\textit{N}\} words. Start with ``Abstract:".
    \item \textbf{Generate with Subject}: Generate a patent abstract with the subject: \{\textit{Patent Subject}\}. Limit your response to \{\textit{N}\} words. Start with ``Abstract:".
\end{itemize}

\subsubsection{Poem}
\begin{itemize}
    \item \textbf{Polish:} Help me polish the following poem: \{\textit{Poem}\}. Limit your response to \{\textit{N}\} words. Start with ``Poem:".
     \item \textbf{Generate with Summary}: Generate a poem with the following summary: \{\textit{Poem Summary}\}. Limit your response to \{\textit{N}\} words. Start with ``Poem:".
    \item \textbf{Generate with Title}: Generate a poem with the title: \{\textit{Poem Title}\}. Limit your response to \{\textit{N}\} words. Start with ``Poem:".
\end{itemize}

\begin{figure*}
    \centering
    \includegraphics[width=1\linewidth]{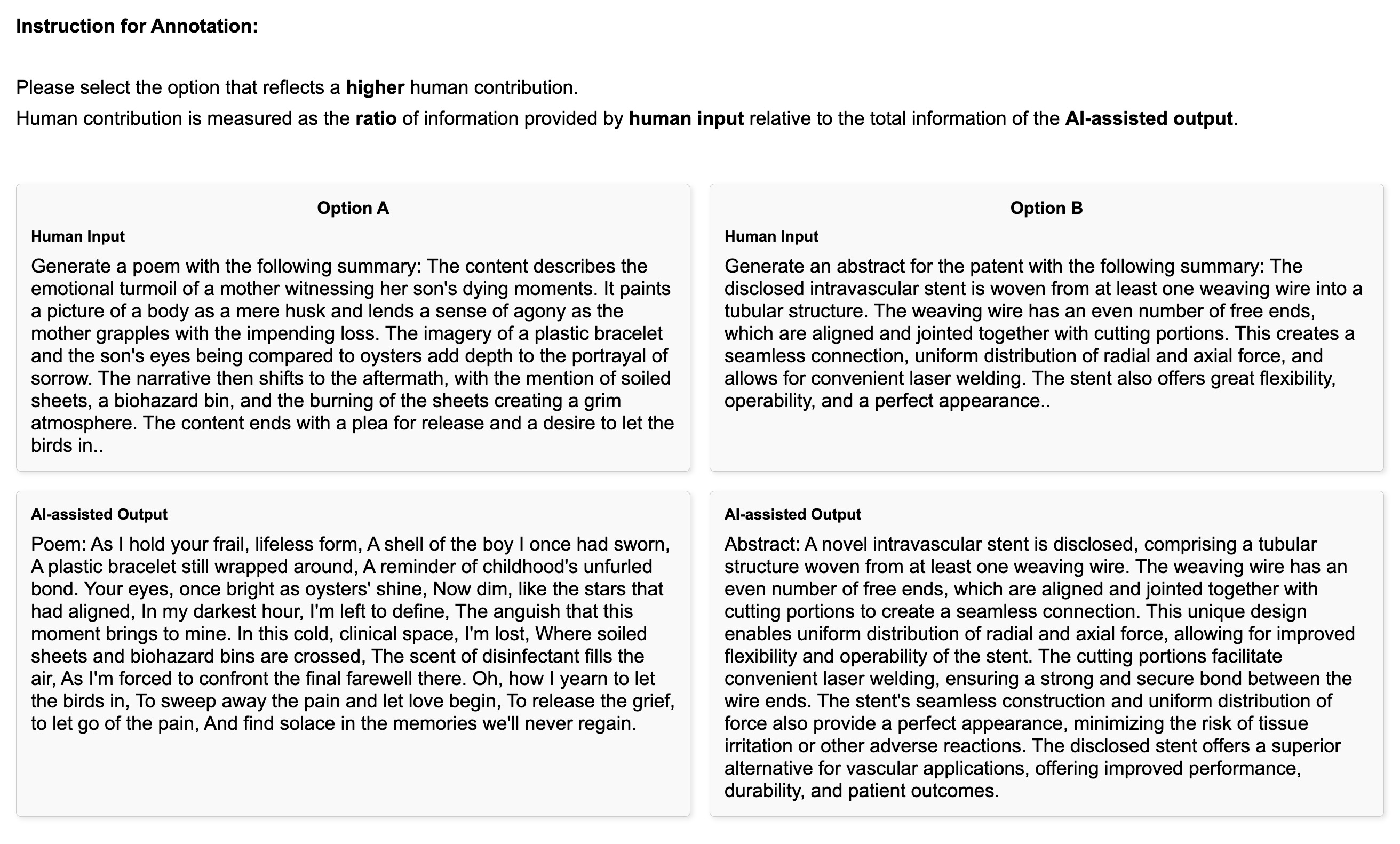}
    \caption{\re{Illustration of the annotation interface.} }
    \label{fig:annotation}
\end{figure*}
\begin{figure*}
    \centering
    \includegraphics[width=1\linewidth]{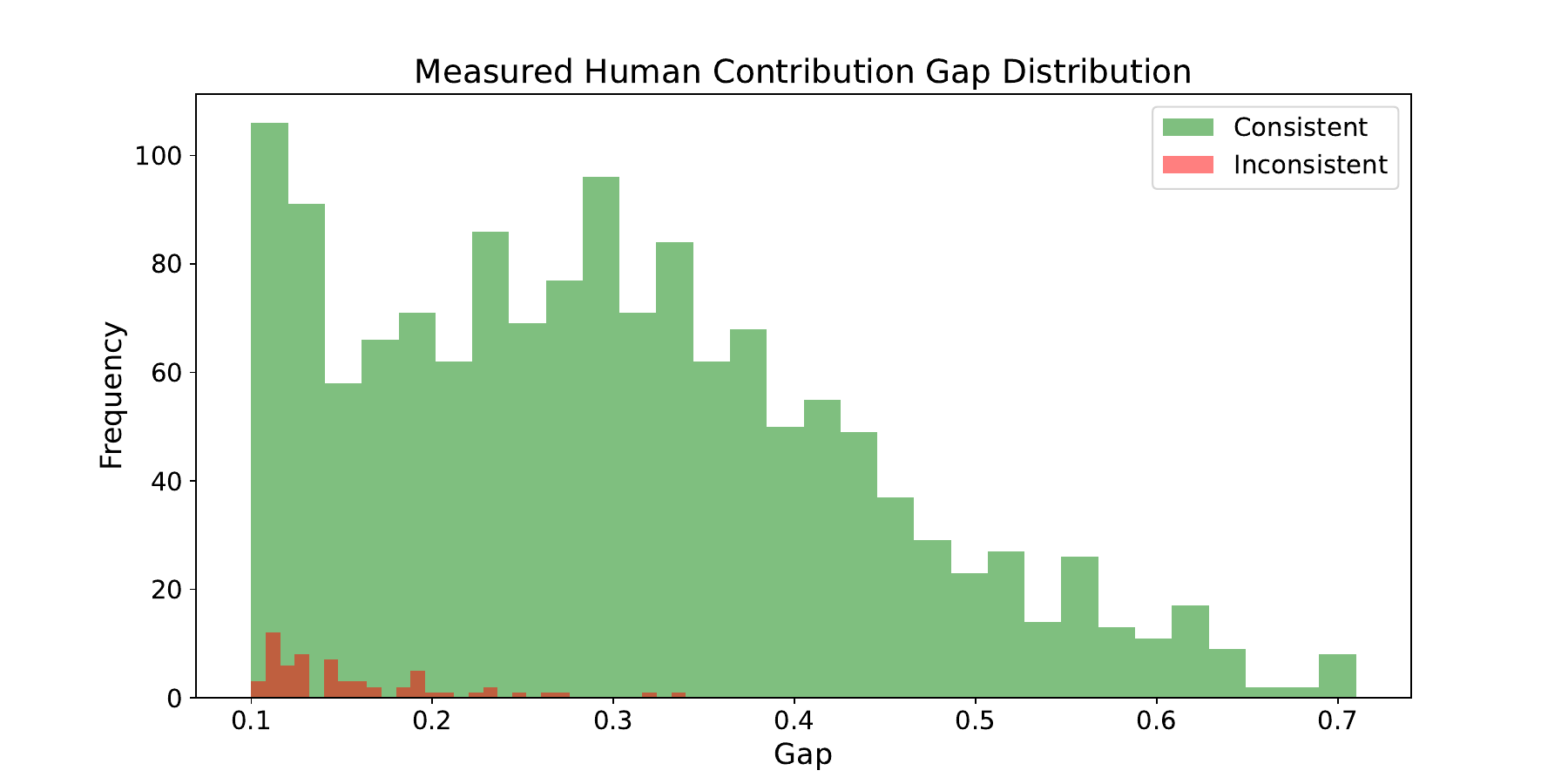}
    \caption{\re{Distribution of measured human contribution gaps for data consistent and inconsistent with human annotations.} }
    \label{fig:gap}
\end{figure*}
\subsection{Model Setups}
\label{supp-modelsetup}
In this section, we detail the versions and parameters of the models we apply in the experiments as follows. 

\textbf{API Settings:} For evaluation, we use Llama-3 (\texttt{Llama-3.1-8B-Instruct}) and Mixtral (\texttt{Mixtral-8x7B-v0.1}). 
The generation models further include GPT-3.5 (\texttt{gpt-3.5-turbo-0125}), GPT-4o (\texttt{gpt-4o-2024-08-06}), \re{Gemini-2.0 Flash (\texttt{gemini-2.0-flash}), and Claude-3.5 Haiku (\texttt{claude-3-5-haiku-20241022})}.

\re{\textbf{Model Selection Rationale:} The rationale for the choice of models in our experiments is as follows. In the main experiments (Figure~\ref{fig:self} and \ref{fig:estimation}), we employed two state-of-the-art open-source language models: Llama-3 and Mixtral.
In the experiment on generalization of our method (Figure~\ref{fig:other-news}), we explored various combinations of generation models and surrogate testing models. 
For the ablation and additional exploratory experiments, we used Llama-3 as the default model to ensure consistency across different settings.
}

\subsection{Human Annotation}
\label{supp-human}
For the human-annotated dataset, we select \re{$1,500$} pairs of input-output data points generated by Llama-3 across all four domains and categories, ensuring that the measured human contribution gap between the two points exceeds 0.1. This threshold is chosen to enhance the meaningfulness of the comparisons and minimize ambiguity during evaluation.

\re{We conducted the annotation using the Labelbox platform. Three platform labeling service annotators, proficient English speakers experienced in expert-level image and text labeling, were recruited for two weeks (80 hours). Each of the 1,500 data points was independently annotated by all three annotators. The final label for each instance was determined by majority voting. The inter-annotator agreement rate—defined as the proportion of cases where all three annotators provided the same label—is $91.47\%$. Figure~\ref{fig:annotation} illustrates the annotation interface used in this process.
}

\new{Figure~\ref{fig:gap} shows the distribution of the measured human contribution gaps, as determined by our proposed method, for data consistent and inconsistent with human annotations. The results indicate that most of the inconsistent cases are associated with smaller measured gaps, further demonstrating the alignment of our measurement with human annotations.}

\subsection{Hyperparameter Setups}
\label{supp-hyper}
In the estimation framework, we define a threshold $\tau$, which serves as a lower bound for 
the geometric mean of the conditional probabilities $p_\theta(\boldsymbol{y} \mid \boldsymbol{x}) ^{1/N}$ if input $\boldsymbol{x}$ could plausibly
generate output $\boldsymbol{y}$ using model $M_\theta$. Correspondingly,  $-\log\tau$ serves as the upper bound for the mean conditional information (or surprisal) $-\log p_\theta(\boldsymbol{y} \mid \boldsymbol{x})/N$ of generating $\boldsymbol{y}$ given $\boldsymbol{x}$. There are several considerations for selecting the hyperparameter threshold $\tau$ used in the estimation framework.

\re{First, the threshold $\tau$ is inherently tied to the generative distribution of the model in use. For models with highly concentrated generative distributions—that is, those that assign high probability to a small subset of plausible outputs—a relatively high threshold $\tau$  for the geometric mean of the conditional probabilities $\left(p_\theta(\boldsymbol{y} \mid \boldsymbol{x}) \right)^{1/N}$ is appropriate, which corresponds to a relatively low threshold $-\log\tau$ for the mean conditional information $-\log p_\theta(\boldsymbol{y} \mid \boldsymbol{x})/N$. This is because, in such models, plausible generations are expected to have relatively high conditional probabilities and thus lower conditional information values. Conversely, for models with more diffuse or uniform generative distributions, plausible outputs may be assigned lower probabilities on average. In such cases, a lower threshold $\tau$ (i.e., a higher threshold $-\log \tau$) is more appropriate.}

\re{Second, ideally,  we would calibrate $\tau$ using a large amount of empirical data from the target domain and model, consisting of paired examples where human input $\boldsymbol{x}$ generates AI-assisted output $\boldsymbol{y}$. This would allow us to observe the distribution of the geometric mean conditional probabilities  (or equivalently, the average conditional information), and to set threshold $\tau$ accordingly to reflect a meaningful boundary between plausible and implausible generations within the testing domain. However, in practice, such domain-specific calibration may increase the barrier to use. Therefore, we provide a general, model-specific reference value for $\tau$ that serves as a reasonable default for real-world applications. Users can still adjust $\tau$ based on their specific domain or sensitivity requirements.}

\re{Third, since it is infeasible to obtain the full generative space of a model for selecting a general threshold value, we approximate it using samples from WildChat, a real-world AI-assisted generation dataset that also serves as out-of-distribution (OOD) data in the evaluation on our constructed dataset. Specifically, we use the set of $2,500$ prompts as those sampled in the Applications to Real-World AI-Assisted Generation section, generate five outputs for each prompt using the target models (Llama-3 and Mixtral), and compute the distribution of average conditional information for each model. We then set $-\log \tau$ to the empirical mean plus one standard deviation of this distribution, representing a reasonable upper bound for the mean conditional information  $-\log p_\theta(\boldsymbol{y} \mid \boldsymbol{x})/N$ on the plausible generation range. This results in  $-\log \tau=0.5496$ for Llama-3 and $0.4155$ for Mixtral in our experiments.  This approach provides a general, model-specific reference value for the threshold, serving as a practical default for real-world applications. Users may still adjust it based on domain-specific requirements or sensitivity preferences.}

\section{Method: Estimating Human Contribution in AI-Assisted Generation Without Human Input}
\label{supp-method}
\new{
In many scenarios requiring the evaluation of human contribution, the generation process of AI-assisted output  $\boldsymbol{y}$ can be complex (e.g., involving iterative back-and-forth refinements) or entirely unavailable. To address this challenge, we extend our framework by introducing the human contribution estimation problem and a corresponding solution that is applicable to any text output, without requiring prior knowledge of the human input  $\boldsymbol{x}$ or the generation process.}

\new{
\textbf{Problem Definition:}
First, we define a research problem aimed at evaluating human contribution without reliance on $\boldsymbol{x}$. It is noteworthy that the same output $\boldsymbol{y}$ can result from various human inputs $\boldsymbol{x}$, which are associated with differing levels of human contribution.
For instance, an AI-generated abstract based solely on a subject (representing minimal human contribution) could also be produced through repetition, where human contribution approaches 100\%.
To establish a well-defined problem, we frame the research challenge as determining the \textit{minimum human contribution} required for human input $\boldsymbol{x}$ to plausibly generate AI-assisted output $\boldsymbol{y}$.
Specifically, this involves quantifying the proportion of information content within the AI-assisted output that is \textit{necessarily} provided by humans.}

\new{
Our focus on minimal human contribution stems from several practical considerations:
First, all outputs can theoretically be generated from a prompt with nearly 100\% human contribution, such as prompting LLMs for repetition. Estimating minimal human contribution provides a reference range, emphasizing the lower bound of human-contributed information content required to plausibly generate the output.
Second, in the context of widespread LLM adoption, minimal human contribution evaluates the \textit{unique original informativeness value} provided by humans—content that LLMs cannot easily generate. 
This distinction is critical in various applications, such as curating high-value data for training next-generation models, assessing creative or original works that reflect human ingenuity and effort, and identifying unique contributions in academic peer reviews. 
}
% \re{Despite the practical utility of this reframed research objective, we emphasize that due to the absence of human input $\boldsymbol{x}$, the outcome no longer reflects the exact human contribution for the given output $\boldsymbol{y}$, but instead serves as a referential lower bound.}

\new{\begin{definition}[Minimal Human Contribution]
Given an AI model $M_\theta$ and the AI-assisted generated content $\boldsymbol{y}$, the minimal human contribution $\hat{\phi}$ is defined as the ratio of the minimal mutual information $\hat{I}(\hat{\boldsymbol{x}}, \boldsymbol{y})$—where $\hat{\boldsymbol{x}}$ represents the minimally informative human input required to plausibly generate $\boldsymbol{y}$—to the self-information $I(\boldsymbol{y})$:
$\hat{\phi} = \frac{\hat{I}(\hat{\boldsymbol{x}}; \boldsymbol{y})}{I(\boldsymbol{y})}.$
\end{definition}}

\new{\textbf{Solution:} 
To solve this problem, we first need to establish a reasonable criterion for determining whether $\boldsymbol{x}$ could plausibly generate $\boldsymbol{y}$.
Considering the generative process of an autoregressive LLM, each token $y_i$ is sampled from a generative distribution $p_\theta(y_i \mid \boldsymbol{y_{<i}}, \boldsymbol{x})$ conditioned on the previous tokens and $\boldsymbol{x}$. 
For $\boldsymbol{y}$ to be considered plausibly generated by $\boldsymbol{x}$, we require that the geometric mean of the conditional probabilities $p_\theta(y_i \mid \boldsymbol{y}_{<i}, \boldsymbol{x})$ across all tokens exceeds a predefined threshold $\tau$:
\begin{equation}
    \left( \prod_{i=1}^{N} p_\theta(y_i \mid \boldsymbol{y}_{<i}, \boldsymbol{x}) \right)^{1/N} > \tau,
\end{equation}
where $N$ is the number of tokens in $\boldsymbol{y}$.  
This condition ensures that, overall, the likelihood of $\boldsymbol{y}$ being sampled from the distribution of model $M_\theta$ given input $\boldsymbol{x}$ at each step is sufficiently large.
It prevents any token in $\boldsymbol{y}$ from having an excessively low probability of being sampled, maintaining the credibility of $\boldsymbol{x}$ as the source of $\boldsymbol{y}$. 
\re{The rationale and empirical values for the threshold $\tau$ are provided in section~\ref{supp-hyper}.}
% The threshold $\tau$ can be chosen based on the characteristics of the generative distribution of evaluated LLM as well as the requirements to balance between permissive and strict criteria. 
% In the evaluation for the paper, $\tau$ is set as $0.65$ for Llama-3 and $0.7$ for Mixtral according to their observed distribution in generation.
}

\new{Based on this criterion, we can estimate the minimum human contribution $\hat{\phi} = \frac{\hat{I}(\hat{\boldsymbol{x}}; \boldsymbol{y})}{I(\boldsymbol{y})}$. The denominator $I(\boldsymbol{y})$ can be directly computed; thus, our primary task is to compute a \textit{lower bound} for the mutual information $I(\boldsymbol{x}, \boldsymbol{y}) = I(\boldsymbol{y}) - I(\boldsymbol{y} \mid \boldsymbol{x})$ by deriving an \textit{upper bound} for the conditional self-information $I(\boldsymbol{y} \mid \boldsymbol{x})$.}

\new{According to our plausibility criterion, we have:
\begin{equation}
    p_\theta(\boldsymbol{y} \mid \boldsymbol{x}) = \prod_{i=1}^{N} p_\theta(y_i \mid \boldsymbol{y}_{<i}, \boldsymbol{x}) > \tau^N,
\end{equation}
where $N$ is the total number of tokens in $\boldsymbol{y}$, and $\tau$ is the predefined plausibility threshold, as detailed in section~\ref{supp-hyper}.
From this, the upper bound for the conditional self-information $I(\boldsymbol{y} \mid \boldsymbol{x})$ can be computed as:
\begin{equation}
    I(\boldsymbol{y} \mid \boldsymbol{x}) = -\log p_\theta(\boldsymbol{y} \mid \boldsymbol{x}) < -N \log \tau.
\end{equation}
This can be interpreted as follows: when $\boldsymbol{x}$ plausibly generates $\boldsymbol{y}$, the likelihood of $\boldsymbol{y}$ being generated based on $\boldsymbol{x}$ is sufficiently high, implying that the conditional information (or surprisal) of generating $\boldsymbol{y}$ given $\boldsymbol{x}$ should be bounded. From this, we derive the mutual information lower bound:
\begin{equation}
    I(\boldsymbol{x}; \boldsymbol{y}) \geq I(\boldsymbol{y}) + N \log \tau.
\end{equation}}
\new{Finally, we can compute the estimated minimal human contribution as:
\begin{equation}
\label{est}
\hat{\phi} = \frac{\hat{I}(\hat{\boldsymbol{x}}; \boldsymbol{y})}{I(\boldsymbol{y})} = \frac{I(\boldsymbol{y}) + N \log \tau}{I(\boldsymbol{y})}.
\end{equation}
This framework enables the estimation of minimal human contribution $\hat{\phi}$ without requiring knowledge of $\boldsymbol{x}$. 
It can be broadly applied to real-world evaluations of human contribution for any given text, providing a practical solution for scenarios where the AI-assisted generation process is unknown.
}

\begin{figure*}[t!]
    \centering
    \includegraphics[width=1.0\linewidth]{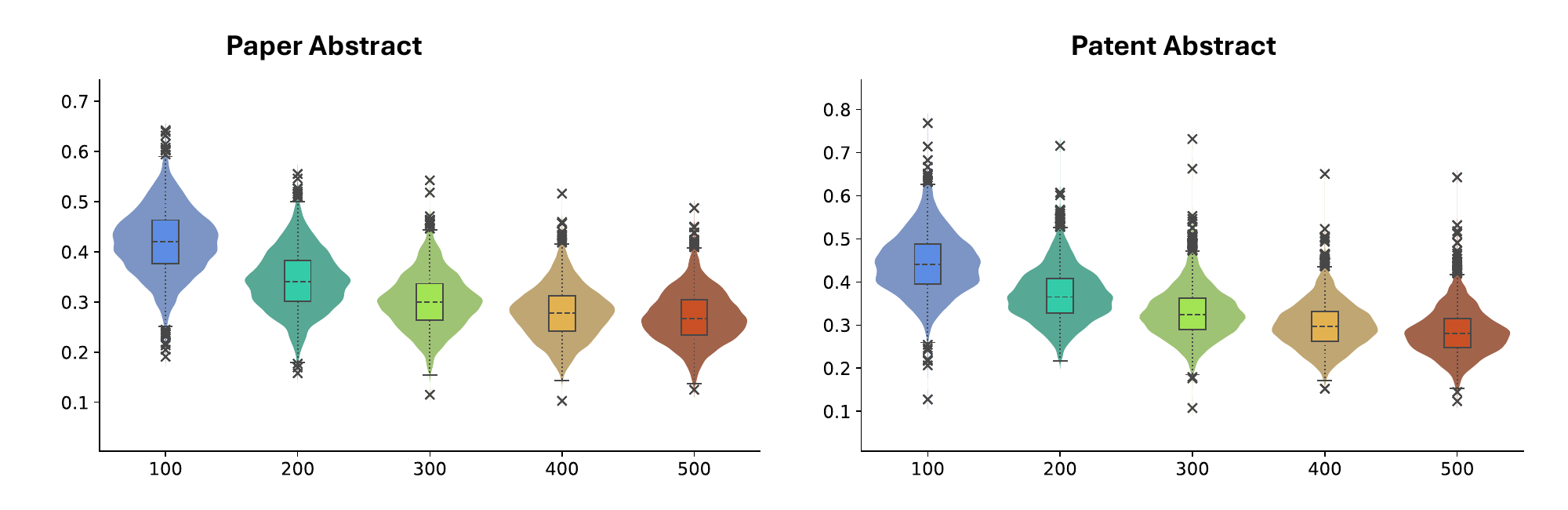}
    \includegraphics[width=1.0\linewidth]{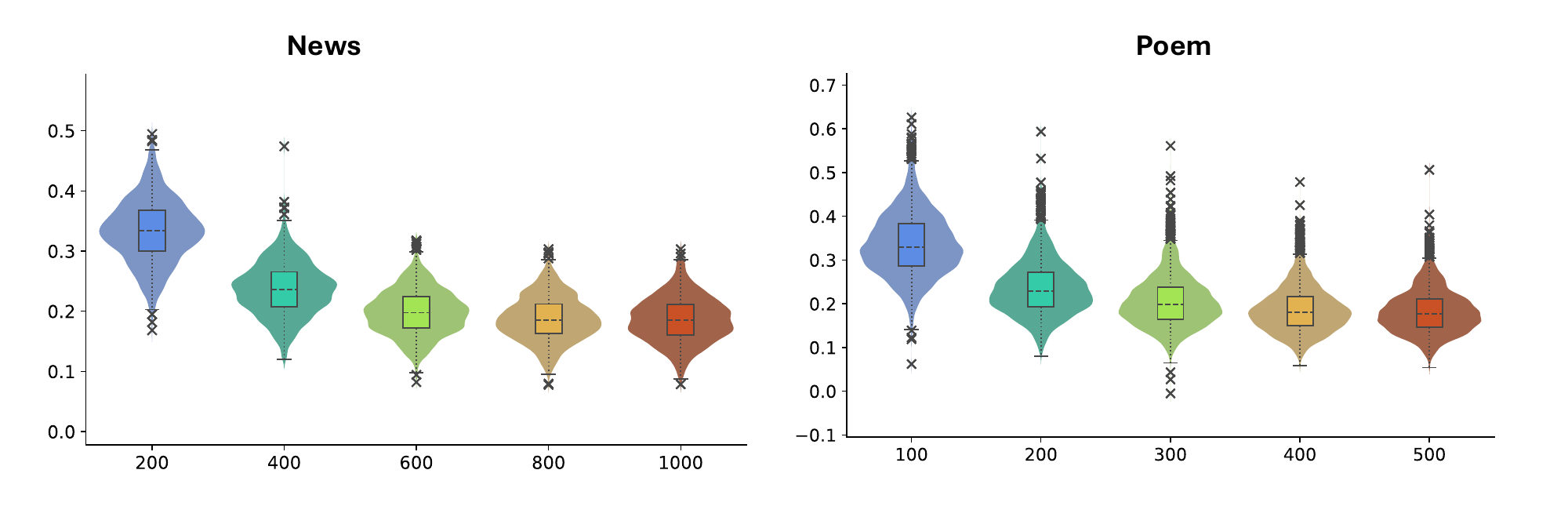}
    \caption{
The distribution of the proposed measure for AI-assisted generations of varying lengths, generated with \textbf{identical titles} using Llama-3.  Overall, the results are consistent with our expectation that, given the same amount of human input information, longer AI-assisted outputs exhibit smaller measured human contributions.
} 
    \label{fig:length}
\end{figure*}
\section{Additional Experimental Results}

\subsection{Impact of Content Length}
\label{supp-length}
In addition to the varying levels of human contribution present in our constructed dataset, we further validate our method by manipulating the total information content of the output while controlling the information content of the input.
By doing so, we aim to verify whether our measurement exhibits the expected property that, when the input is controlled, increasing the output information leads to a lower human contribution (as defined by the ratio of human-contributed information relative to the total information in the output).

To achieve this, we adjust the output length requirements in the input prompts while keeping the other input information constant. We assume that much longer outputs generally have higher information content. Specifically, we use Llama-3 to \textbf{generate AI-assisted outputs of varying lengths from the identical titles} by specifying the desired output length in the prompt.
% varying the length of the AI-assisted content that is generated. 
% This helps us determine whether our method adequately evaluates the proportion of human contribution in AI-assisted generated content when the same human input information yields AI-assisted outputs of different lengths.
% Intuitively, when human informational input remains constant, the longer the generated content, the smaller the measured human contribution should be. To verify this, we use Llama-3 to generate AI-assisted outputs of varying lengths from titles by specifying the length of the AI-assisted output in the prompt.

The results, as shown in Figure~\ref{fig:length}, align with our expectations: 
\new{as the AI-assisted output contains more information content, the proportion of human informational contribution relative to the total informational content decreases, as reflected by our measurement of human contribution.}
% as we require AI-assisted output to \new{ moew information conetnt},  human informational contribution relative to  total informational content decreases, as does our measurement of human contribution.

\begin{figure*}[t!]
    \centering
    \includegraphics[width=0.8\linewidth]{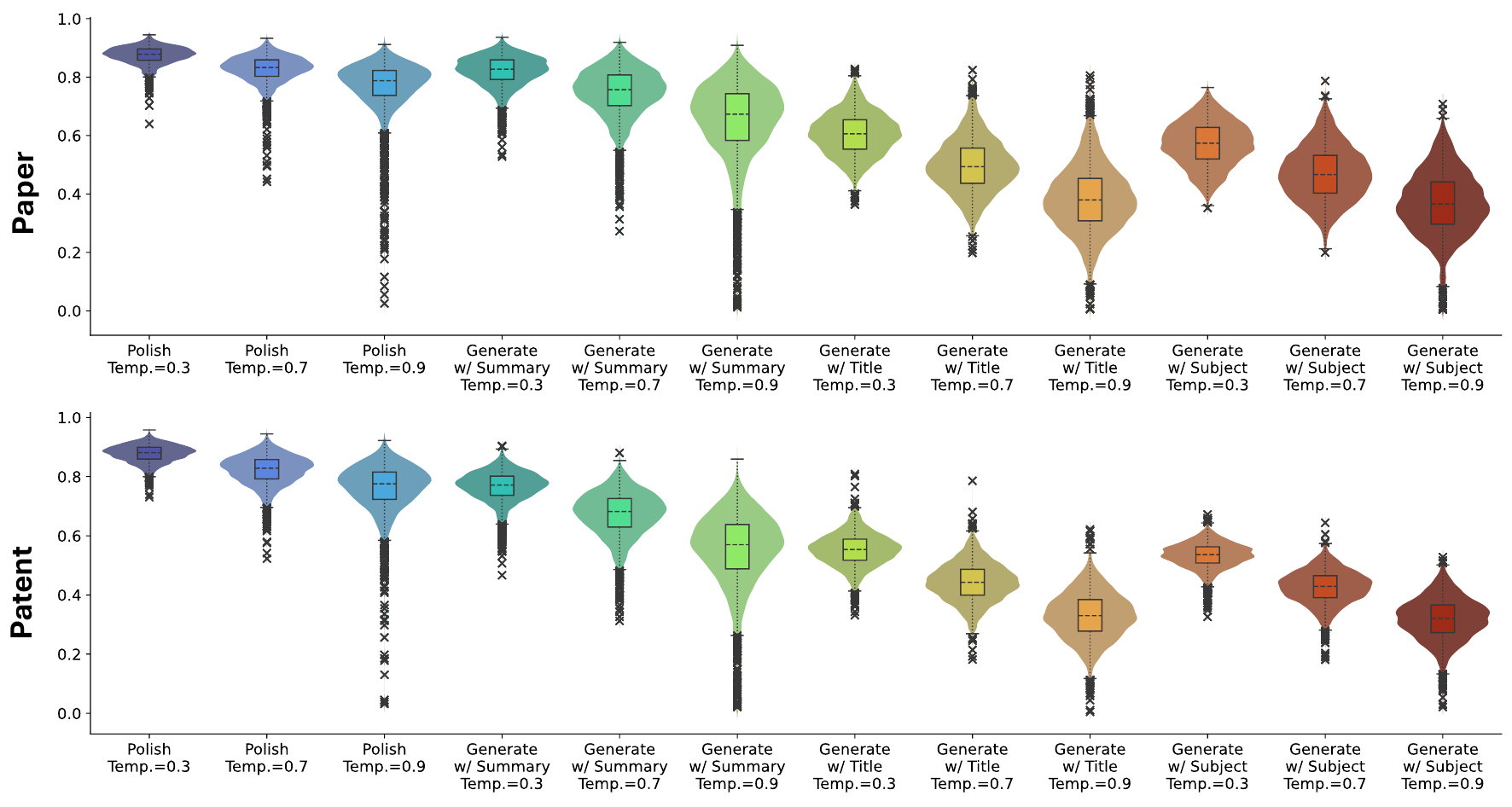}
        \includegraphics[width=0.8\linewidth]{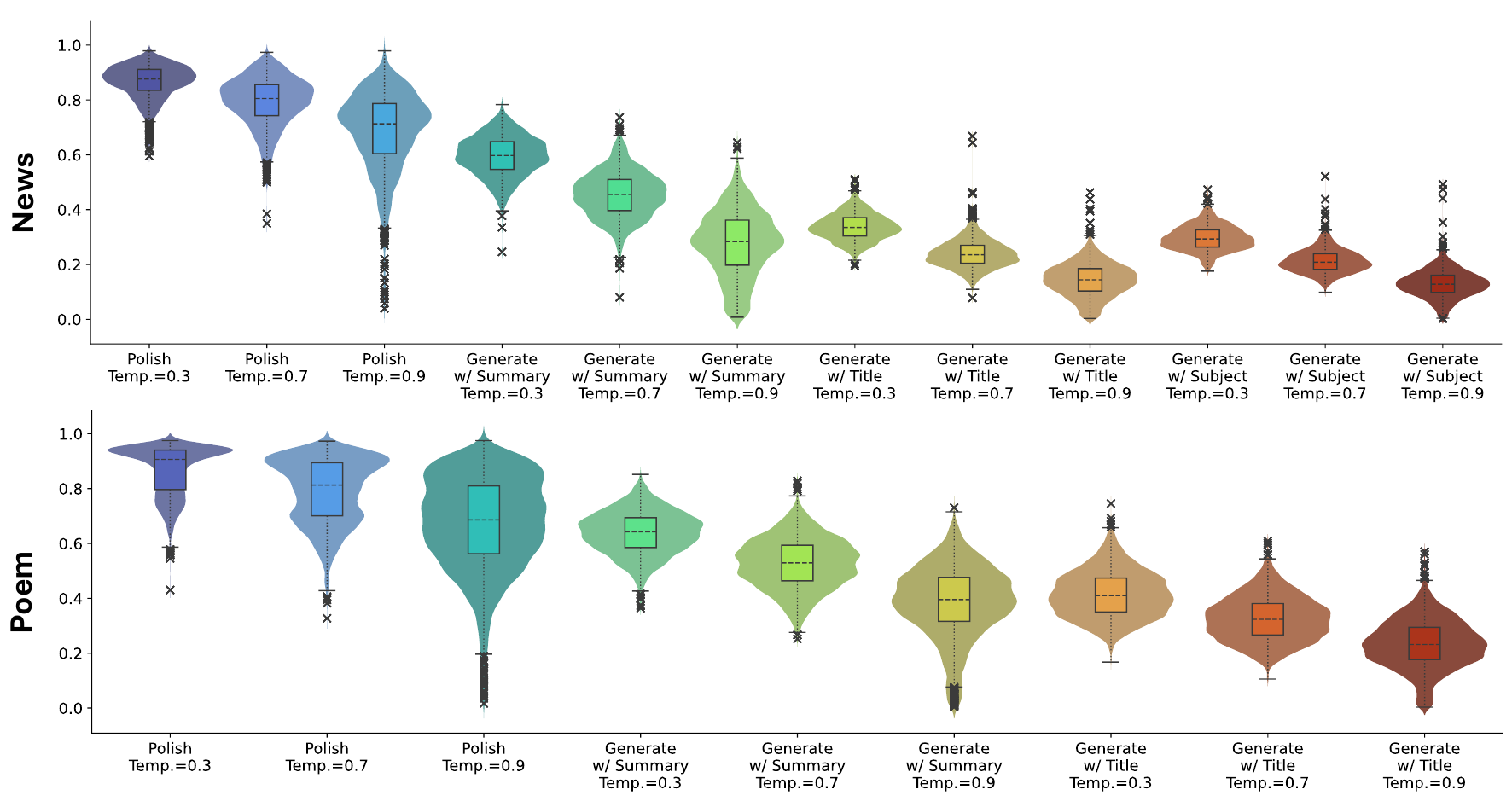}
    \caption{
\new{The distribution of the proposed measurement outcomes for AI-assisted generation under different sampling temperature settings using Llama-3. Overall, the results are consistent with our expectation that higher sampling temperatures lead to smaller measured human contributions.}
} 
    \label{fig:temp}
\end{figure*}

\begin{figure*}[t!]
    \centering
    \includegraphics[width=0.8\linewidth]{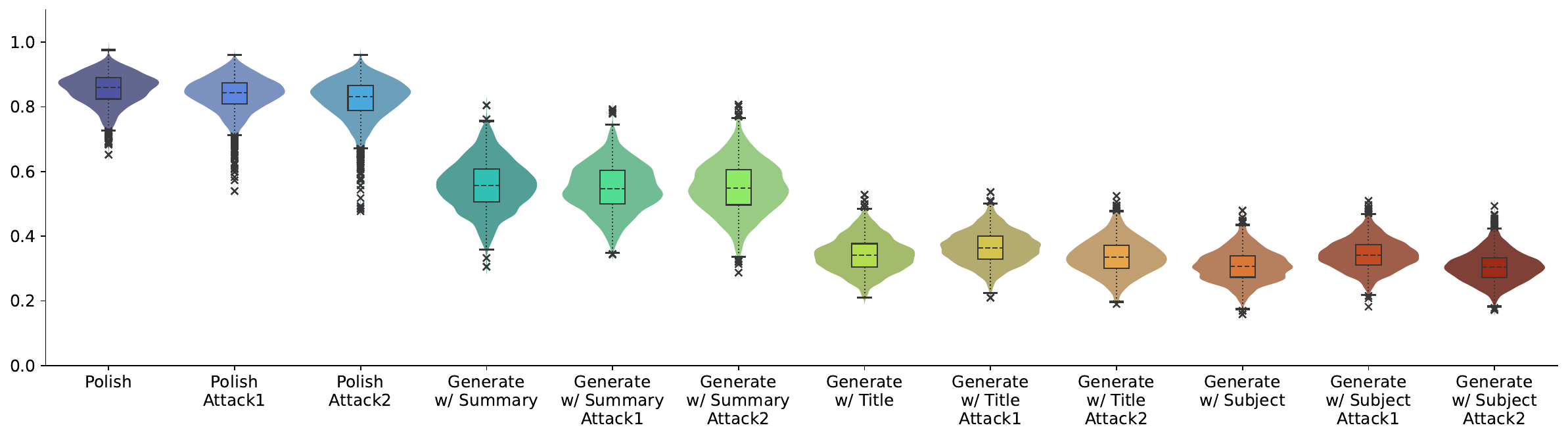}
        \includegraphics[width=0.8\linewidth]{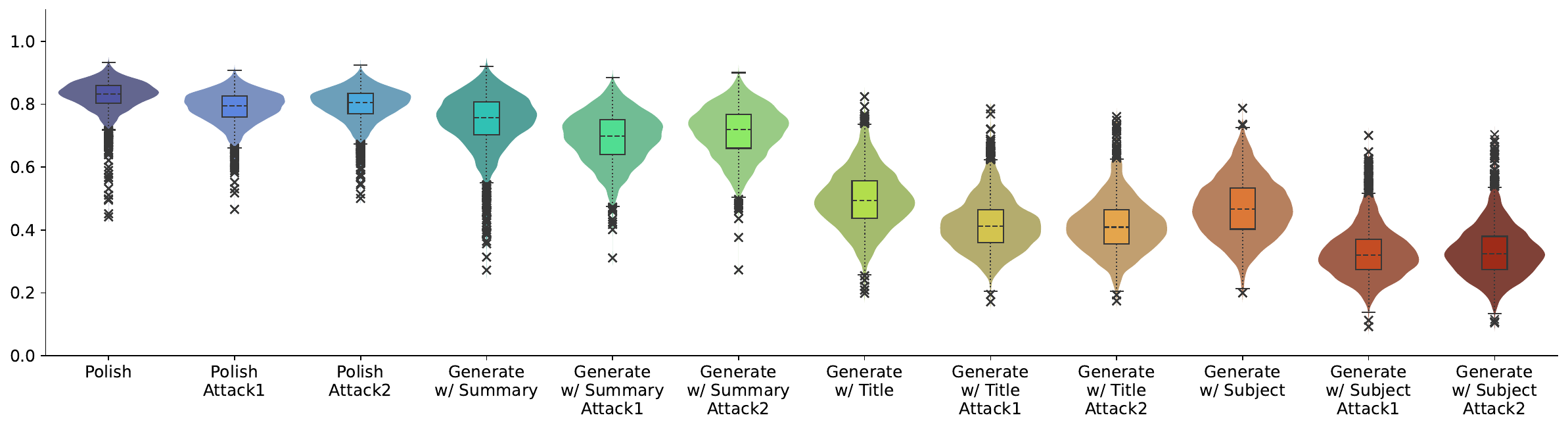}
            \includegraphics[width=0.8\linewidth]{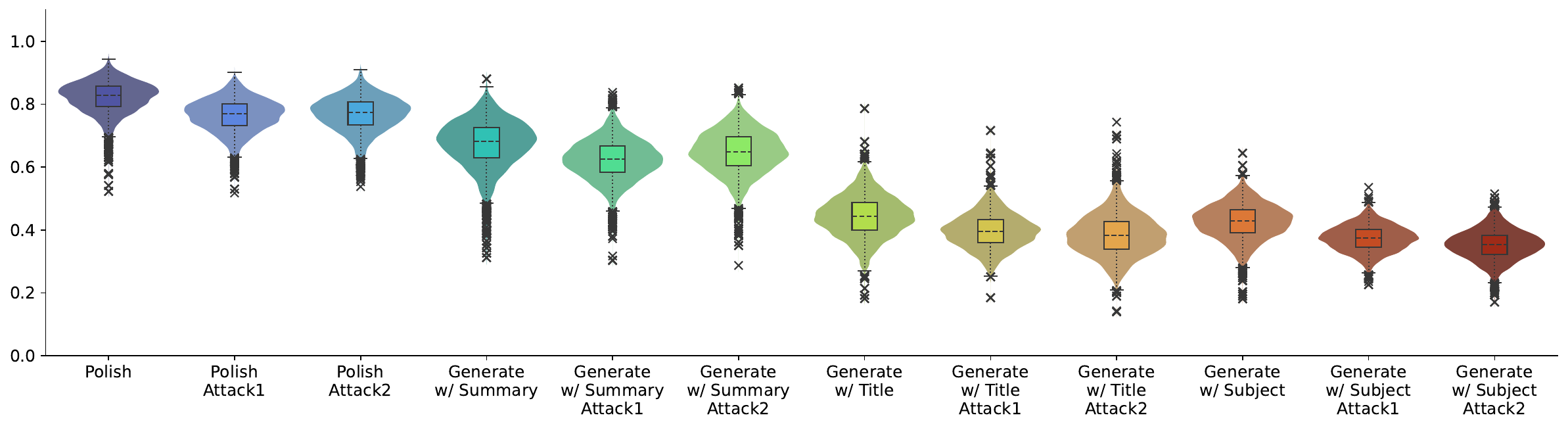}
                \includegraphics[width=0.8\linewidth]{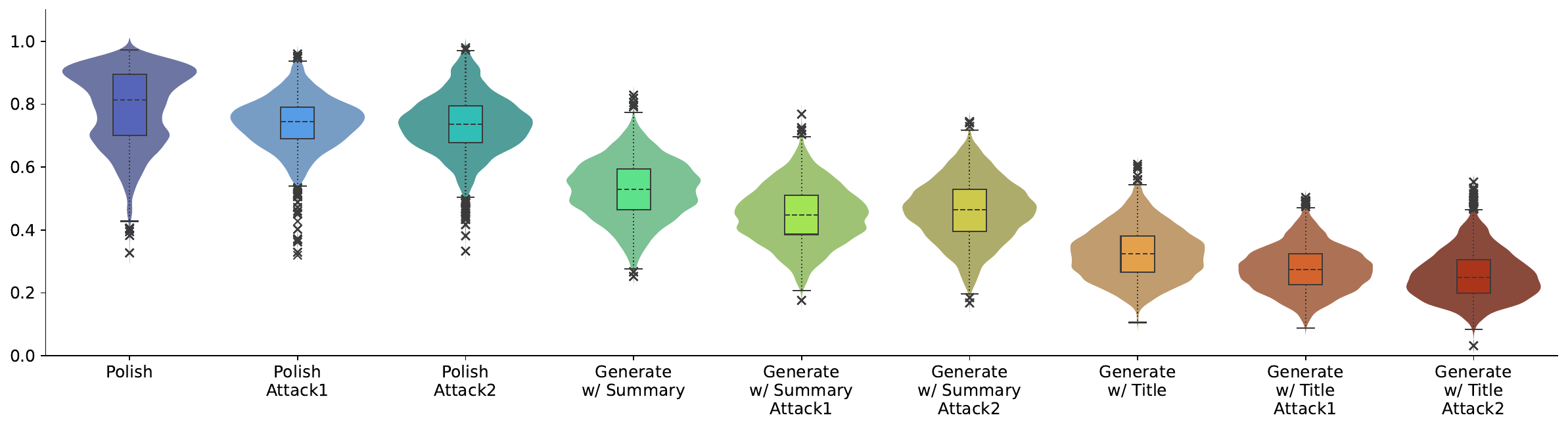}
    \caption{
 The distribution of the outcomes of the proposed measure for the constructed dataset with and without adaptive attacks using Llama-3. Overall, the adaptive attacks have little to no influence on the measurement outcomes.
} 
    \label{fig:adapt-news}
\end{figure*}
\subsection{Impact of Generative Model Temperature}
\label{supp-temp}
\new{Temperature refers to the parameter that controls the randomness of output in a generative model. During the generation process, a lower temperature makes the model more deterministic by favoring high-probability tokens, while a higher temperature introduces greater variability, resulting in more diverse but less predictable outputs.
In this section, we study the impact of generative model temperature on the outcomes of our measurement.
While the ``human effort” (human input) remains unchanged across different temperature settings, we expect human contribution—defined as the proportion of informational content in the output attributable to human input—to decrease as the temperature increases. This is because the output becomes less dependent on the input information and increasingly reflects the model’s explorative generation.
In extreme cases, when the temperature approaches infinity, the output of the language model would become entirely independent of the human input, resulting in an expected human contribution of zero. 
}

\new{
Figure~\ref{fig:temp} demonstrates the comparison between temperature settings from 0.3 to 0.9 during the generation with Llama-3.
We can observe that generally, a higher temperature leads to a smaller measured human contribution across different generation categories. This validates that our measurement accurately reflects the diminishing influence of human input on AI-assisted output with increasing temperature of the generative model.}

\subsection{Resilience to Adaptive Attacks}
\label{supp-adaptive}
We further investigate whether adaptive attacks could be employed in real-world applications to artificially inflate measured human contribution. 
\re{To this end, we design two adaptive attacks. For each attack, we append an instruction to the original input prompts that does not provide additional information but may influence the AI’s generation process in a way that potentially increases the measured human contribution. Specifically, the two instructions used are: (1) “Always choose words you rarely use.” and (2) “Mimic human writing.”
}
The first instruction influences the model's generation probabilities to produce less frequently used words, thereby attempting to increase the perceived information content (surprisal). The second instruction guides the model to generate text that closely resembles human writing, thereby attempting to increase the perceived human contribution. \re{}

Figure~\ref{fig:adapt-news} shows the results of our measure with and without attacks using Llama-3.
We can observe that our measure remains robust against the adaptive attacks. This aligns with our expectation, as we measure human contribution by utilizing the ratio of the mutual information between human input and AI-assisted output to self-information of the AI-assisted output itself. These non-informational instructions for manipulating the output do not significantly affect our measure.

\begin{figure*}
    \centering
    \includegraphics[width=1.0\linewidth]{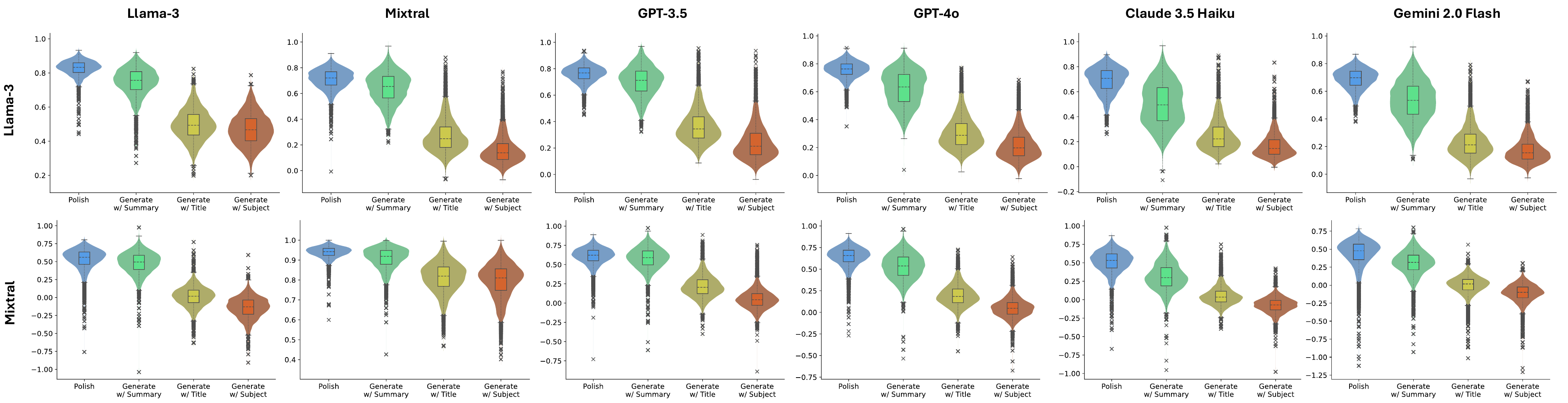}
    \caption{
 The distribution of the proposed measurement on the constructed dataset of paper abstracts for various generation models (columns) and surrogate models for measurement (rows). Overall, for each model pair, the proposed measure exhibits the desired trend that lower measured values are obtained for the generated content with less human contribution.
} 
    \label{fig:other-paper}
\end{figure*}
\begin{figure*}
    \centering
    \includegraphics[width=1.0\linewidth]{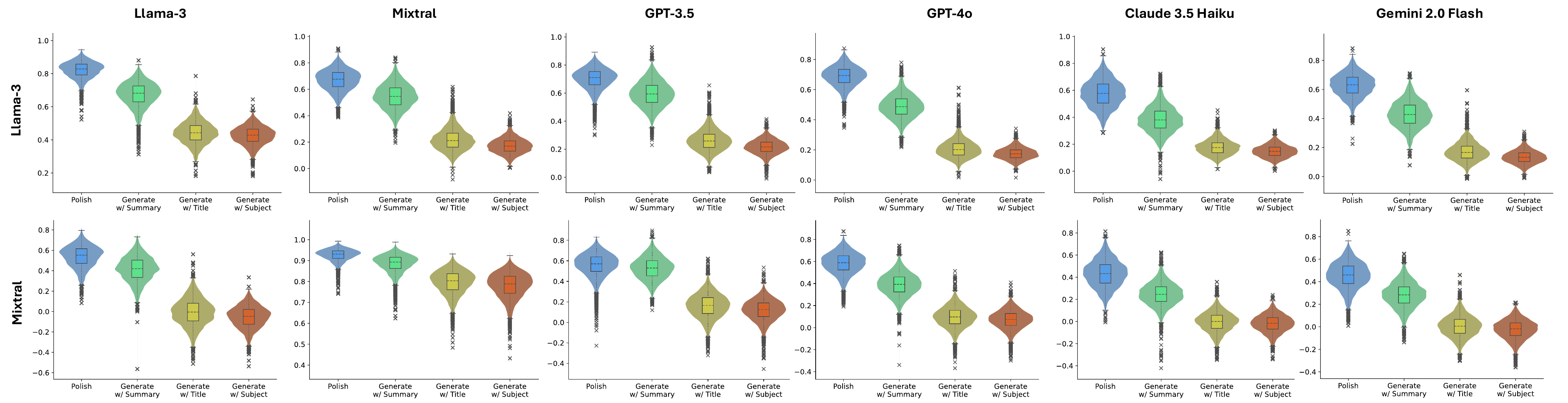}
    \caption{
 The distribution of the proposed measurement on the constructed dataset of patent abstracts for various generation models (columns) and surrogate models for measurement (rows). Overall, for each model pair, the proposed measure exhibits the desired trend that lower measured values are obtained for the generated content with less human contribution.
} 
    \label{fig:other-patent}
\end{figure*}
\begin{figure*}
    \centering
    \includegraphics[width=1.0\linewidth]{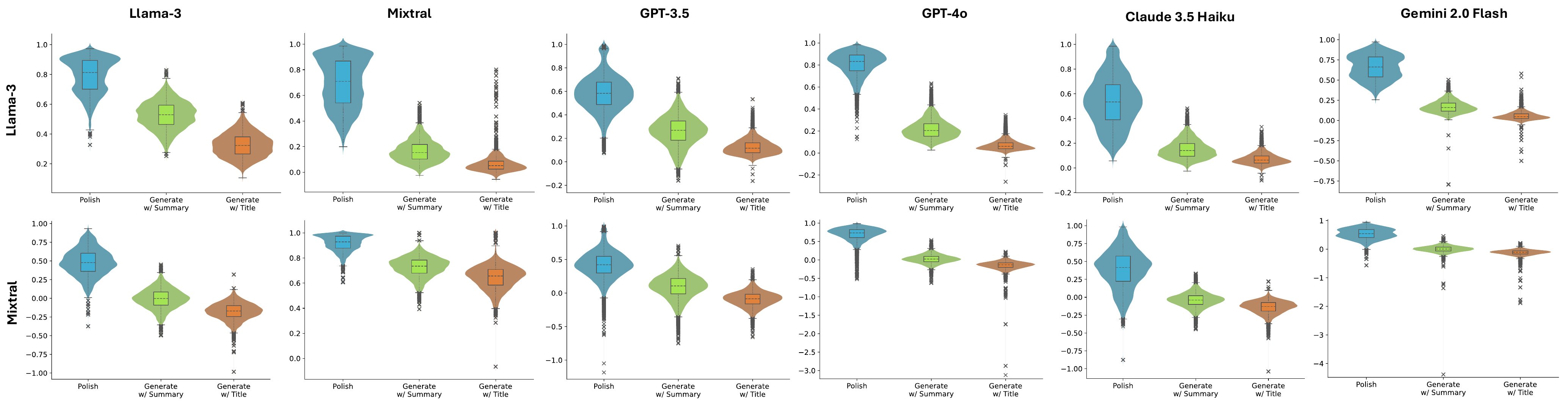}
    \caption{
 The distribution of the proposed measurement on the constructed dataset of poems for various generation models (columns) and surrogate models for measurement (rows). Overall, for each model pair, the proposed measure exhibits the desired trend that lower measured values are obtained for the generated content with less human contribution.
} 
    \label{fig:other-poem}
\end{figure*}

\subsection{Generalization of Our Method}
\label{supp-generalization}
Besides the generalization results on news data presented in the main paper, we further show the results on paper abstract, patent abstract, and poem in Figure~\ref{fig:other-paper}, \ref{fig:other-patent}, and \ref{fig:other-poem}, respectively.

\begin{figure*}[t!]
    \centering
    \includegraphics[width=1\linewidth]{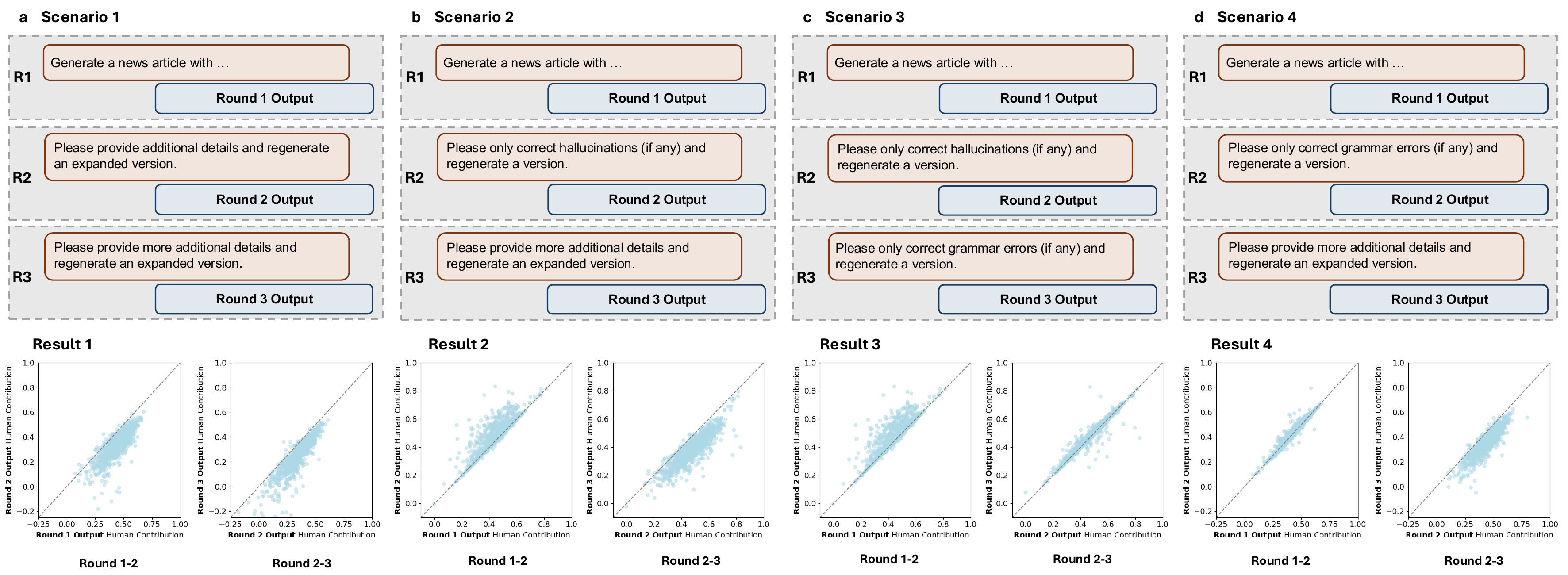}
    \caption{
 \new{\re{Illustration of the four multi-round generation scenarios and corresponding evaluated results. The scatter plots (Results 1–4) visualize pairwise comparisons of estimated human contribution between rounds. For each subplot, the x-axis and y-axis represent estimated human contribution values for two consecutive rounds. The diagonal line indicates perfect agreement.}} } 
    \label{fig:multi}
\end{figure*}

\subsection{Multi-Round Generation Analysis}
\label{supp-multi}
In addition, we further investigate the validity of our evaluation in the context of multi-round
\re{generation. As illustrated in Figure~\ref{fig:multi}, we test four commonly used multi-round generation scenarios and examine whether the estimated human contribution values exhibit expected patterns and variations across outputs from different rounds.
Specifically, we sample 1,000 news articles from the original dataset for the first round of generation using Llama-3, covering all generation modes. We then introduce second and third rounds of interactions, involving three common real-world operations: adding details, correcting hallucinations, and correcting grammatical errors. 
Human contribution values for outputs from different rounds are computed using our output-only estimation framework, and pairwise comparisons are conducted across consecutive rounds to analyze how contribution levels evolve in response to each operation. }

% \re{Intuitively, when the model is prompted to add more details, the proportion of human contribution is expected to decrease, as the model generates additional content with minimal new input from the human.
% For polishing and grammar correction, human contribution is expected to remain relatively stable, as these operations typically involve minor, surface-level edits.
% In contrast, when correcting hallucinations, human contribution may increase if the output from the previous round contains hallucinated content. Since hallucinations often diverge from the original input or factual context, correcting them results in the output becoming more grounded in the human input, thereby increasing the proportional informational contribution of the human in the final content.
% If no hallucinations are present, the estimated contribution is expected to remain stable.}

\begin{table*}[!t]
\centering
\small
\resizebox{\textwidth}{!}{%
\begin{tabular}{lcccc}
\toprule
& CHEAT~\cite{yu2025cheat} & HC3-English~\cite{guo2023close} & HC3-Chinese~\cite{guo2023close} & Ghostbuster~\cite{verma2024ghostbuster} \\
\midrule
AUC & 0.9593 & 0.9931 & 0.9710 & 0.9857 \\
\bottomrule
\end{tabular}%
}
\caption{ROC--AUC performance of the proposed estimation framework on AI-generated content detection benchmarks.}
\label{tab:detection}
\end{table*}
\re{The experimental results are presented in Figure~\ref{fig:multi}. For each scenario, the left panel shows the relationship between the estimated human contribution for the Round 1 output (x-axis) and the Round 2 output (y-axis), while the right panel depicts the relationship between the Round 2 output (x-axis) and the Round 3 output (y-axis). These scatter plots visualize pairwise differences in estimated human contribution across consecutive rounds. The diagonal line indicates perfect agreement between rounds.
The results generally align with our expectations. First, we observe consistency across rounds: the human contribution differences introduced by various generation modes (e.g., polishing, generation from a summary, title, or subject) in the first round tend to persist in subsequent rounds.
Second, for the “adding details” operation (Rounds 1–2 and 2–3 in Scenario 1; Rounds 2–3 in Scenarios 2 and 4), most points lie below the diagonal, indicating that human contribution generally decreases as the model adds more content with minimal new human input.
For “grammar correction” (Round 2-3 in Scenario 3 and Round 1–2 in Scenario 4), the results are the most stable, as grammar errors in model-generated text are generally sparse, leading to minimal changes in estimated human contribution. 
For the “correcting hallucinations” operation (Rounds 1–2 in Scenarios 2 and 3), many points appear above the diagonal, suggesting that human contribution can increase when hallucinated content from the previous round is corrected and the output becomes more evidently rooted in the human-provided information. If no hallucinations are present, the contribution remains stable.}

% Case studies are demonstrated in section~\ref{supp-casemulti}.}

\subsection{AI-generated Content Detection}
\label{supp:detection}
Our estimation methods proposed in section~\ref{sec:estimation} can also be applied to AI-generated content detection tasks. To evaluate this capability, we conduct experiments on four detection benchmarks spanning multiple domains, including academic abstracts, student essays, creative fiction, news articles, Reddit posts, Wikipedia entries, as well as medical and financial texts. The evaluation covers two languages, English and Chinese.

For each dataset, we randomly sample 1{,}000 human-written texts and 1{,}000 AI-generated texts. Using only the text, we compute the human-contribution estimation score and assess its discriminative performance using ROC–AUC.

The results (Table~\ref{tab:detection}) show consistently high AUC values across datasets, indicating that the output-only estimation framework can effectively distinguish AI-generated content from human-written content. This further demonstrates that the proposed output-only estimation metric captures meaningful signals related to the degree of human involvement.

\end{document}